# Balloon Flight Test of a Compton Telescope Based on Scintillators with Silicon Photomultiplier Readouts

P. F. Bloser[1], J. S. Legere, C. M. Bancroft, J. M. Ryan, M. L. McConnell

Space Science Center, University of New Hampshire, Durham, NH 03824, USA

## Abstract

We present the results of the first high-altitude balloon flight test of a concept for an advanced Compton telescope making use of modern scintillator materials with silicon photomultiplier (SiPM) readouts. There is a need in the fields of high-energy astronomy and solar physics for new medium-energy gamma-ray (~0.4 – 10 MeV) detectors capable of making sensitive observations of both line and continuum sources over a wide dynamic range. A fast scintillator-based Compton telescope with SiPM readouts is a promising solution to this instrumentation challenge, since the fast response of the scintillators permits both the rejection of background via time-of-flight (ToF) discrimination and the ability to operate at high count rates. The Solar Compton Telescope (SolCompT) prototype presented here was designed to demonstrate stable performance of this technology under balloon-flight conditions. The SolCompT instrument was a simple two-element Compton telescope, consisting of an approximately one-inch cylindrical stilbene crystal for a scattering detector and a one-inch cubic $LaBr_3$:Ce crystal for a calorimeter detector. Both scintillator detectors were read out by 2 × 2 arrays of Hamamatsu S11828-3344 MPPC devices. Custom front-end electronics provided optimum signal rise time and linearity, and custom power supplies automatically adjusted the SiPM bias voltage to compensate for temperature-induced gain variations. A tagged calibration source, consisting of ~240 nCi of $^{60}$Co embedded in plastic scintillator, was placed in the field of view and provided a known source of gamma rays to measure in flight. The SolCompT balloon payload was launched on 24 August 2014 from Fort Sumner, NM, and spent ~3.75 hours at a float altitude of ~123,000 feet. The instrument performed well throughout the flight. After correcting for small (~10%) residual gain variations, we measured an in-flight ToF resolution of ~760 ps (FWHM). Advanced scintillators with SiPM readouts continue to show great promise for future gamma-ray instruments.

---

[1] Corresponding author:
Address: University of New Hampshire, 8 College Road, Durham, NH 03824, USA
Phone: +1 603 862 0289
Fax: +1 603 862 3584
Email: Peter.Bloser@unh.edu

# 1. Introduction

The fields of high-energy astronomy and solar physics are in need of new instrumentation to make sensitive observations of medium-energy (~0.4 – 10 MeV) gamma rays from space. In this difficult energy range the instrument of choice for high sensitivity is the Compton telescope, since the coincidence requirement and coarse imaging ability greatly suppress background from cosmic ray interactions in the instrument and spacecraft. The prototypical Compton telescope, and the only one to perform sensitive observations in space, was the COMPTEL instrument on board the Compton Gamma Ray Observatory (CGRO) mission [1]. COMPTEL was composed of two detector layers separated by 1.6 m: a scattering layer (D1), made of up organic liquid scintillator tanks, and a calorimeter layer (D2), made up of inorganic NaI(Tl) scintillator crystals. Scintillation light produced by gamma-ray interactions in the detector volumes (ideally Compton scattering in D1 followed by photoelectric absorption in D2) was recorded by photomultiplier tubes (PMTs), which permitted the location and energy deposit of each interaction to be measured. In addition, the pulse shape of the D1 signals, and the time difference between the D1 and D2 triggers, were recorded. Only events for which the pulse shape indicated a gamma-ray interaction, and the time difference was consistent with a downward-moving scattered photon (~5 ns), were accepted; these pulse-shape discrimination (PSD) and time-of-flight (ToF) discrimination methods proved crucial for rejecting background and achieving a useful signal-to-noise ratio.

In the years since the demise of CGRO in 2000, most efforts to advance MeV gamma-ray science have been based on instruments composed of semiconductor detectors with very good (≲ 1%) energy resolution, largely motivated by the goal of studying gamma-ray lines produced in cosmic nucleosynthesis and solar flares. The most prominent example in astronomy is the European INTEGRAL mission [2], which features a cryogenically cooled Ge spectrometer array in the SPI instrument [3] and an array of small CdTe detectors in the IBIS instrument [4]. Despite many notable scientific successes at hard X-ray energies, however, INTEGRAL has been unable to significantly improve on COMPTEL's sensitivity at energies ≳ 1 MeV. This is partly because the telescope concept used in both instruments, a coded-aperture imager with massive shielding surrounding the detector arrays, is not optimal for high sensitivity to medium-energy gamma rays, especially in terms of background rejection. Indeed, it has long been recognized that a next-generation Compton telescope would be required to succeed COMPTEL, even before the launch of INTEGRAL, and much work has been devoted to the study of different technological approaches (e.g., [5] and references therein). The most advanced such concept, the Compton Spectrometer and Imager (COSI) balloon instrument [6], also employs cooled Ge as the detector material in order to emphasize sensitivity to lines. Like INTEGRAL, however, Compton telescopes based on semiconductor detector materials suffer from inherent drawbacks that limit their utility for MeV gamma-ray measurements. These include small detector volumes, merely moderate efficiency to gamma ray interactions in the case of Ge, and slow response times that preclude the use of ToF discrimination for event reconstruction and background rejection. We note that, despite its excellent energy and position resolution, the COSI instrument requires an active shield made of heavy inorganic scintillator in order to achieve a low background [6].

In solar physics the most prominent mission dedicated to gamma ray measurements in recent years has been RHESSI [7], which features a Ge spectrometer [8] paired with rotation modulation collimators for spectroscopy and imaging of narrow lines from solar flares. As with INTEGRAL in astronomy, however, the success of RHESSI above ~1 MeV has been limited: despite highly significant scientific contributions from the imaging of hard X-rays and the 2.2 MeV neutron capture line [9], only a handful of gamma-ray line flares have been studied in detail above 2.2 MeV. The most comprehensive set of such flares remains that obtained by the NaI(Tl) gamma-ray spectrometer (GRS) on the Solar Maximum Mission [10]. One reason for this is, once again, the limited stopping power of Ge to MeV gamma rays. Another reason is the slow response time of the semiconductor detectors, which produces pileup and dead time effects during bright flares. Finally, the background in the (unshielded) RHESSI spectrometer is quite high, limiting sensitivity to gamma rays from weaker flares. Indeed, the faintest solar flare known to emit nuclear line radiation was detected by COMPTEL [11], once again demonstrating the superior sensitivity of Compton telescopes in this energy range despite only moderate energy resolution.

The examples of INTEGRAL and RHESSI indicate that sensitive measurements of medium-energy gamma rays from space over a wide dynamic range are best accomplished using large volumes of detectors with good stopping power and fast time response. As exemplified by COMPTEL [1], scintillator materials read out by PMTs have a long history of filling this role, both in astronomy (e.g., [12 – 16]) and in solar physics (e.g., [17, 18]). Scintillator materials have the additional advantages of room temperature operation, radiation hardness, and relatively low cost. When configured as a Compton telescope, the fast time response of scintillators permits the use of the proven ToF discrimination technique for event reconstruction and background rejection. All of these historical advantages have been further strengthened in recent years by the development and commercial availability of newer, higher-performance scintillators. Organic liquids, such as the NE213 used in the COMPTEL D1 layer, are toxic and difficult to handle, while solid plastics have low light output. Recently, however, improved growth techniques [19] have made solid organic crystal scintillators, such as stilbene and p-terphenyl, commercially available in large volumes. These crystals feature improved light output compared to liquids, and retain the ability to distinguish between gamma-ray and neutron interaction signals, due to their differing light decay times, via the PSD technique. Stilbene has been flown in space in the LEND instrument on the Lunar Reconnaissance Orbiter [20]. On the inorganic side, the most promising new scintillator is $LaBr_3$:Ce [21], though other materials such as $CeBr_3$ [22] are of interest as well. Compared to the NaI(Tl) used in the COMPTEL D2 layer, $LaBr_3$:Ce offers greatly improved stopping power, energy resolution ($\lesssim$ 3% FWHM at 662 keV, vs. ~7%), and time response (26 ns decay time, vs. 230 ns). We note that, since most cosmic gamma-ray lines (e.g., from supernovae and solar flares) are expected to be Doppler broadened by ~2 – 3%, the energy resolution of these modern advanced scintillators is sufficient for the majority of applications. $LaBr_3$:Ce (hereafter $LaBr_3$) has been the subject of study for a variety of future astrophysics and space science mission concepts (e.g., [23, 24]).

All scintillator-based instruments are limited by the mass, volume, and power of the scintillation light readout device. Historically this device has been the PMT, which, despite providing high gain (~$10^6$) and fast time response (~ns), is fragile, bulky, and requires high voltage (~1000 V), all of which are drawbacks for a space-based instrument. In addition, it is especially important for a Compton telescope to minimize passive material near the sensitive detector elements, both because such material can become activated by cosmic ray interactions and produce background, and because incident gamma rays that scatter in passive components before interacting in a detector will not be reconstructed properly. For these reasons, fully realizing the potential of advanced scintillators for a next-generation Compton telescope will require replacing PMTs with compact, low-voltage light sensors that retain equivalent quantum efficiency, gain, and timing performance. Fortunately, silicon photomultipliers (SiPMs) are now commercially available that promise to meet this need. Originally developed for high-energy physics applications [25], SiPMs are undergoing rapid development for use in conjunction with various scintillators in nuclear medicine detector systems [26]. SiPMs consist of two-dimensional arrays of small (~50 μm) avalanche photodiode elements, or "cells," that are operated in limited Geiger mode and read out in parallel. The summed output signal is proportional to the total number of cells that are triggered by the absorption of an optical photon. When coupled to a scintillator, a SiPM will therefore measure the brightness of the optical pulse generated by a gamma-ray interaction. This combination of proportional response with Geiger-mode avalanche operation results in a solid-state light detector with gain (~$10^6$), speed (several ns), and photon detection efficiency (~30%) nearly equivalent to a PMT. Several SiPMs that have become commercially available in recent years feature quantum efficiency well matched to the blue (~380 – 420 nm) scintillation light of materials such as stilbene and $LaBr_3$.

The Space Science Center at the University of New Hampshire (UNH) has for several years been investigating the use of advanced scintillators and SiPMs in instrumentation for high-energy astronomy and solar physics. This work has included the study of $LaBr_3$ for coded-aperture telescopes [27]; the laboratory investigation of SiPM readouts for scintillator spectrometers [28 – 31]; a successful balloon flight test of a Fast Compton Telescope (FACTEL) prototype with a $LaBr_3$ D2 layer and ~1 ns ToF resolution [32]; the successful balloon flight test of a simple $LaBr_3$/SiPM spectrometer [33]; and a simulation study of the potential sensitivity of an advanced scintillator Compton telescope with SiPM readouts on an Explorer-class space mission [34]. In this paper we describe our most sophisticated experimental effort to date: the balloon flight test of a simple scintillator Compton telescope prototype with SiPM readout. The Solar Compton Telescope (SolCompT) experiment consisted of a stilbene D1 and $LaBr_3$ D2, both read out by SiPM arrays, with a tagged $^{60}$Co source in the field of view to provide a known gamma-ray source to measure. We describe the instrument and balloon payload in detail, and present the flight data analysis and results, which demonstrate stable performance and good ToF resolution suitable for future high-energy astronomy and solar physics instruments.

**2. Experimental Setup: Instrument and Balloon Payload**

*2.1 SolCompT Instrument*

The SolCompT instrument was a two-element Compton telescope designed to demonstrate stable performance of a multi-component, SiPM-based instrument under balloon-flight conditions. The D1 and D2 scintillator elements, together with their associated SiPM light readouts, are shown in Fig. 1. The D1 detector (Fig. 1, right) was a cylindrical stilbene crystal, 3 cm in diameter and 3.8 cm tall, wrapped in several layers of white Teflon tape and an outer layer of black electrical tape. This crystal was originally purchased from Proteus, Inc., and tested with SiPM readout for pulse height and PSD performance as part of the Dose Spectra from Energetic particles and Neutrons (DoSEN) project at UNH [30, 35]. The D2 detector (Fig. 1, left) was a 26 mm × 26 mm × 26 mm cubic LaBr$_3$ crystal purchased from Saint-Gobain. This crystal and its surrounding white reflective material were sealed in a hermetic aluminum package with a 1-mm thick optical light guide on one face.

The two scintillators were each read out by a 2 × 2 array (26 mm × 26 mm total light collecting area) of Hamamatsu S11828-3344 MPPC devices mounted on a custom SiPM detector readout board (Fig. 1, bottom). This board, also originally designed for DoSEN [30, 35], sums the 16 anode signals of each individual S11828-3344 array to provide four outputs, one for each "quadrant," into 50 Ω coax cables (Fig. 2a). When working with such large-area, large-capacitance SiPMs, it is very important that the readout provide a very low input impedance in order to maintain a fast rise time and avoid non-linearities in the pulse height response [30, 31, 36]. However, under these conditions the impedance due to parasitic inductance (of leads, etc.) can begin to dominate, and this inductance will interact with the large detector capacitance to produce resonance effects. We therefore used a transformer to match the 50 Ω coax cable to the characteristic resonant impedance (~5 Ω) due to detector capacitance and parasitic inductance. The transformer turns ratio was chosen so as to minimize the resonance (the impedance across a transformer scales as the square of the turns ratio). This approach is illustrated in Fig. 2a as a transformer (in an auto-transformer configuration) at the summed anode of each S11828-3344 device. Any residual resonance of the detector capacity (~5000 pF) and lead and transformer leakage inductance (1 – 2 μH) is damped by a series resistor at each summed anode. Using this transformer readout, the rise time for each summed quadrant was ~11 – 12 ns for the stilbene D1, and ~18 – 19 ns for the LaBr$_3$ D2, as measured with a Tektronix TDS 3024B digital oscilloscope. Using a charge preamp with a 47 Ω series resistor, as suggested in early Hamamatsu documentation for small (~1 mm) MPPCs, the rise time of a summed S11828-3344 was ~300 ns [30].

For the SolCompT instrument configuration the four quadrant signals from each detector board were roughly scaled to account for inherent gain variations between the different S11828-3344 devices (factors of ~2) and then summed in a high-speed buffer amplifier on a separate board (Fig. 2b). The buffer amplifier provided a gain of 1.8 into a 50 Ω load, with a matched output impedance of 50 Ω to minimize high-speed reflections. This second board (Fig. 2b) also provided a bias voltage (~71 V) that was automatically scaled to compensate for temperature-induced gain variations in the SiPMs. A stable, low-noise voltage was generated with a precision sine wave inverter design (UNH patented) with no harmonic content in the detectors' frequency range and ripple < 1 mV peak-to-peak. A 10 kΩ thermistor bead was mounted on the detector board to monitor the temperature of the SiPMs. A "temperature conditioner" on the power supply board converted the thermistor resistance into a linear voltage proportional to the temperature, with an absolute accuracy of ±1ºC between 5ºC and 40ºC. The output of the temperature conditioner was then scaled and summed with the input supply to provide the appropriate gain compensation for the SiPMs. The scaling factor was set using a potentiometer; the expected value for the S11828-3344 was ~55 mV/ºC, as is typical for Hamamatsu MPPC devices [31]. Scaled voltages representing the temperature, scaled bias voltage, and detector current were available on the power supply board to be read as housekeeping data (see Sec. 2.3).

The D1 and D2 scintillators were optically coupled to the SiPM arrays using Dow Corning Sylgard 3-6636 Silicone Dielectric Gel and secured with black electrical tape. The two detector boards were then mechanically mounted using standoffs to form a two-element Compton telescope with a scintillator center-to-center separation of 15 cm (Fig. 3). This distance was chosen so as to maximize the efficiency for Compton-scatter events by minimizing the detector separation, while still maintaining up-versus-down event discrimination via the expected ~1 ns (FWHM) ToF resolution.

*2.2 Tagged $^{60}$Co Source*

Because the SolCompT instrument was so small, and the balloon flight was only expected to last for a few hours (see Sec. 3.1), a tagged calibration source was constructed to ensure that readily identifiable gamma rays of known energy and location would be available to measure during flight in order to monitor the instrument performance. Three 1-cm cylinders of plastic scintillator doped with $^{60}$Co, each wrapped in white Teflon and black electrical tape, were optically cemented to the face of a 1-inch Hamamatsu R1924A PMT (Fig. 4). Each source had an activity of ~80 nCi, so that the total tagged source strength was ~240 nCi. The front-end electronics generated a fast discriminator pulse when the scintillation light from a beta particle was detected by the PMT, permitting calibration events to be identified.

The tagged source was mounted ~8 cm from the top face of the D1 scintillator at an angle of ~45º from vertical. A simple Monte Carlo simulation using MGGPOD [37] predicted that tagged Compton-scatter events would be detected by SolCompT in this configuration at a rate of ~4 counts per minute.

*2.3 Balloon Payload*

The SolCompT instrument was integrated into a balloon payload consisting primarily of components that had flown twice before, most recently for the FACTEL experiment in 2011 [32], and before that for the Gamma-Ray Polarimeter Experiment (GRAPE) prototype in 2007 [38]. The major elements of the payload are illustrated in Fig. 5, and a system block diagram is shown in Fig. 6.

The SolCompT instrument, power supply/buffer boards, and tagged $^{60}$Co source were mounted near the top of an instrument frame modified from FACTEL. The frame was made of fiberglass-reinforced plastic (FRP) so as to avoid background from the activation of aluminum by cosmic rays in the immediate vicinity of the detectors. The frame was mounted on a round stand made of PVC material to provide thermal isolation. This stand was in turn mounted on the base plate of a small (38 cm diameter × 63 cm tall) cylindrical aluminum pressure vessel with a domed top.

The summed outputs of the D1 and D2 detectors were processed using the same analog channel electronics boards used for FACTEL. These channel boards, one for each layer, contained a fast discriminator, time-to-voltage converters, pulse shaping, and pulse height (PH) track-and-hold circuits. The time-to-voltage circuit on the D1 board provided the ToF value, using the D1 discriminator to generate the START signal, and the D2 discriminator to generate the STOP (see Fig. 6). The full range for this measurement was ±25 ns, giving a hardware coincidence window of 50 ns. The D1 channel board also generated the PSD value for the D1 pulse by comparing the amount of integrated signal in the tail of the pulse to the total. The D1 channel board also contained analog multiplexors that captured and held the temperature, bias voltage, and detector current housekeeping values from the power supply/buffer boards (see Sec. 2.1).

The event readout was controlled by a PIC18F4620 processor chip mounted on a separate processor board. The PIC received an event start signal from either discriminator, digitized all held voltage values (PH, PSD, ToF, and housekeeping) using its internal analog-to-digital converter (ADC), and issued a reset to the channel electronics. The PIC also monitored the discriminator signal (held for 100 ns) from the tagged $^{60}$Co source and generated a flag for each event where this value was high. Different data modes permitted the PIC to record coincident Compton events, D1 single events, D2 single events, or to simply count discriminator pulses within a given time interval in order to calculate trigger rates for D1, D2, or the tagged $^{60}$Co source. The temperature, voltage, and current were read for both D1 and D2 as part of every coincident event. The PIC formatted the data bytes into event words and transmitted them via an RS-232 interface to the serial port of a PC104 single-board computer.

The overall data acquisition was controlled by the PC104 computer, which ran custom software (modified from the successful GRAPE and FACTEL flights) on a Linux platform. The program operated in a continuous loop, sending commands to the PIC processor to acquire coincident Compton events for 5 minutes, followed by 10 s of D1 singles, 10 s of D2 singles, and 10 s of rate calculation for the detectors and the tagged calibration source. These different data types were written to separate data files on the PC104 hard disk and backup solid-state memory. Each individual Compton event received a time stamp from the PC104, whereas the other data types were marked with the time at the beginning of the 10-second acquisition interval. In addition to the detector and tagged source data, the PC104 recorded housekeeping data every ten seconds from a variety of internal sensors (pressure, temperatures, and voltages) using a custom data acquisition board.

Other elements of the balloon payload included two sets of resistive heaters connected to thermal switches to passively maintain an internal temperature of ~20ºC – 25ºC; a low-noise high voltage supply for the R1924A PMT in the tagged calibration source; and a hermetic connector in the base plate of the pressure vessel to permit remote connections to the payload. These included an Ethernet cable to communicate with the PC104 during integration and testing, and connection to an external power supply or the standard balloon-flight battery packs (~28 V). As all science and housekeeping data were recorded on board, no commanding or telemetry were implemented other than two voltages, encoding the internal pressure and temperature, which were available through the connector to be read by the external NASA Consolidated Instrument Package (CIP) and monitored during the flight. This would have permitted the payload to be powered off in the event of a pressure leak so as to avoid damage to the PC104's hard disk.

The assembled payload as flown is shown in Fig. 7 with the pressure vessel top removed.

*2.4 Ground Calibration*

The SolCompT instrument was calibrated in the laboratory at UNH and in the field before flight using radioactive sources. All calibration data were obtained at room temperature, using the full balloon payload readout. The scaling factor for the temperature-compensating bias voltage supply was set to 55 mV/ºC for both D1 and D2; this setting was verified to be approximately correct by taking singles data with each detector over a temperature range of 10ºC – 30ºC in a thermal chamber prior to instrument integration. Due to schedule constraints during final flight campaign preparations, however, the scaling factor could not be finely tuned.

The PH-to-energy calibration for the D1 detector was performed by recording singles spectra from $^{137}$Cs (662 keV) and $^{22}$Na (511 keV and 1275 keV) sources. Since photons at these energies preferentially Compton scatter in the organic stilbene crystal, the energy calibration made use of the Compton edges, as was done previously for FACTEL [32]. A Monte Carlo simulation of the calibration setup was performed, and then an energy-dependent resolution was applied to broaden the simulated Compton edges until they matched the calibration data. This permitted the energies corresponding to the half-maxima of the Compton edges to be identified (496 keV and 1086 keV for the 662 keV and 1275 keV lines, respectively). These two energies were sufficient, since the response of the SiPM to the relatively dim stilbene scintillator is linear over this energy range [30]. The calibrated D1 singles spectra recorded in the field for $^{137}$Cs and $^{22}$Na are shown in Fig. 8, along with the tagged singles spectrum recorded from the $^{60}$Co calibration source.

The PSD response of the D1 detector was calibrated at UNH using a $^{252}$Cf fission source of gamma rays and fast neutrons. The measured PSD parameter is shown as a function of the D1 PH in Fig. 9. Two tracks are evident, due to neutron (upper) and gamma-ray (lower) interactions. Although the separation is not as clean as that obtained previously using laboratory NIM electronics [30], it is sufficient to allow neutrons to be efficiently rejected via a PH-dependent PSD cut above PH values of ~50 (~130 keV) and below ~430 (~1780 keV).

The PH-to-energy calibration for the D2 detector was calculated using Gaussian fits to the 511 keV, 662 keV, and 1275 keV photopeaks recorded in singles mode from $^{137}$Cs and $^{22}$Na sources. A linear fit to the PH vs. energy relation was found to be sufficient over this energy range; that is, the well-known SiPM saturation [31] was not a significant factor due to the relatively large light-collecting area and dynamic range of the SiPM array. The calibrated D2 singles spectra recorded in the field for $^{137}$Cs and $^{22}$Na are shown in Fig. 10. The energy resolution at 662 keV was measured to be 5.30 ± 0.04% (FWHM).

The ToF was calibrated by placing a bright $^{60}$Co source directly between D1 and D2 and recording coincident events from the two simultaneous gamma rays (1173 keV and 1333 keV). Each detector contributes a PH-dependent delay, or offset, to the measured ToF value due to walk in the discriminators on the channel boards. These offsets can be measured and corrected individually for each detector using events with near-constant PH in the other detector, as shown in Fig. 11. Coincident events were first selected for which the D2 PH was within a narrow range near maximum (corresponding to the 1333 keV photopeak). The ToF vs. D1 PH for these events was then plotted, resulting in a narrow band of points (Fig. 11a). The ToF offset relative to some arbitrary value (here channel 327, the ToF value for high PHs in both detectors) was derived from a moving average of the ToF centroid along this band (gray line in Fig. 11a). (Where the bands became steep at small PH values, the fit to the ToF centroid was simply extended by eye.) The procedure was repeated for D2 (Fig. 11b) by selecting events within a narrow range of large D1 PH (corresponding to the 1333 keV Compton edge). Calculating and removing both of these offsets, derived from the D1 and D2 PH values individually, then sharply narrowed the overall ToF distribution, centered on channel ~327.

After applying corrections based on both the D1 and D2 PH values, the channel-to-time calibration was performed by measuring the change in the ToF centroid when calibrated delay cables were introduced into the system. A value of 87.5 ps chan$^{-1}$ was derived. For large PH values in both D1 and D2, the ToF resolution from the coincident $^{60}$Co events was 541 ± 13 ps (FWHM). This resolution is only slightly worse than that measured in the lab between individual elements of the PMT-based FACTEL instrument, albeit at lower energies, after applying a similar PH-dependent correction procedure [32].

Finally, shortly before the balloon flight, a $^{22}$Na source was placed on the outside of the pressure vessel at approximately a 40º angle off the SolCompT telescope axis, and data were recorded using the flight software. Coincident scatter events were analyzed by applying D1 and D2 energy calibrations and ToF calibrations, and then calculating the Compton scatter angle $\varphi$ using the Compton equation

$$\cos\varphi = 1 - \frac{m_e c^2}{E_2} + \frac{m_e c^2}{E_1 + E_2}, \qquad (1)$$

where $E_1$ and $E_2$ are the energy deposits in D1 and D2, respectively, and $m_e c^2$ is the electron rest mass energy (511 keV). The coincident total energy spectrum ($E_{tot} = E_1 + E_2$) for untagged events with ToF > 0 and 25º ≤ $\varphi$ ≤ 45º is shown in Fig. 12. Both gamma-ray lines of $^{22}$Na, 511 keV and 1275 keV, are clearly visible. The energy resolution for the total energy is ~11.2% (FWHM) at 511 keV, and ~5% at 1275 keV.

The event rate for tagged $^{60}$Co events in the integrated payload was ~3 counts per minute, consistent with the expected rate (Sec. 2.2) given imperfect tagging efficiency and uncertainty in the precise activity of the source.

**3. Data Analysis and Results**

*3.1 Balloon Flight*

The SolCompT payload was integrated as a piggyback experiment on a NASA Columbia Scientific Balloon Facility (CSBF) test payload in Fort Sumner, NM. The pressure vessel was installed on the second "deck" of a Long Duration Balloon (LDB) test gondola, in close proximity to the CIP, relay boxes, batteries, and other CSBF equipment (Fig. 13). Battery power (~28 V) was connected through a relay box to permit the payload to be commanded on shortly before launch and off before termination. The internal pressure and temperature were monitored as described in Sec. 2.3.

The LDB test payload carrying SolCompT was launched from Fort Sumner on 24 August 2014 (CSBF flight #651N). The payload achieved a float altitude of ~123,000 feet, and maintained it for ~3.75 hours until termination (Fig. 14). (The relatively short flight duration was due to the desire of CSBF personnel to keep the payload nearby for easy retrieval, once the test of LDB equipment was completed.) After the payload was recovered and returned to Fort Sumner it was confirmed that SolCompT operated normally and recorded data onboard throughout the flight.

Fig. 15 shows the singles count rate in the LaBr$_3$ D2 detector for both untagged (black) and tagged (gray) events. While the tagged event rate remained constant (~6 cts s$^{-1}$), the untagged rate showed the expected behavior for a balloon flight: the rate decreased immediately after launch (shortly before 08:00 Mountain Daylight Time), increased during the ascent to ~200 cts s$^{-1}$ in the Pfotzer Maximum, and then decreased to a constant ~100 cts s$^{-1}$ at float (~10:20 MDT) until termination. These data indicate that the scintillator/SiPM detector combination operated normally throughout the flight, as in previous work [33], and that events from the tagged $^{60}$Co source are correctly identified.

*3.2 Temperature Effects and Corrections*

One of the goals of the SolCompT balloon flight was demonstration of the stability of a SiPM-based instrument to temperature variations. The temperature vs. time recorded throughout the flight by the thermistor bead on the D2 detector board (Sec. 2.1) is shown in Fig. 16. As expected, the temperature varied from ~21ºC – 25ºC at float as the thermal switches cycled the heaters on and off (Sec. 2.3), and reached ~28ºC during the ascent. For comparison, the raw PH vs. time for tagged singles events in the D2 detector is plotted in Fig. 17. The two lines of $^{60}$Co at 1173 keV and 1333 keV are visible at approximately constant PH values between ~500 and 600. Over similar temperature ranges, uncompensated Hamamatsu SiPMs would display gain variations greater than a factor of two [31]. However, residual gain variations at the ~10% level in the SolCompT data are still apparent. This is likely due to the fact that, as described in Sec. 2.4, time constraints prevented careful calibration of the temperature-scaling factor on the power supply board. Another contributing factor is the fact that the thermistor beads used are inherently non-linear. We find a relative change in the reported vs. true temperature of ~0.2ºC over the range 21ºC – 28ºC, which we estimate contributes ~3% gain change.

The tagged calibration source permits a straightforward correction to the detector gains to remove the residual ~10% variation. For D2, the tagged singles PH values corresponding to the 1333 keV line (narrow band of points in Fig. 17 at PH < 600) were smoothed using a moving boxcar average, and a time-dependent scaling factor relative to the maximum value was interpolated for all event times (singles and coincident). A similar procedure was performed for D1 tagged singles using PH values corresponding to the 1333 keV Compton edge. Finally, since the timing delays in each detector depend on the PH, a time-dependent shift in the ToF centroid was fit and interpolated for tagged coincident events in the same way. Thus the tagged $^{60}$Co source permitted all residual temperature effects to be removed from the flight data.

*3.3 Energy Calibration*

After correcting the D1 and D2 PH values for gain variations, it was necessary to derive a new in-flight energy calibration. This was done for each detector using singles events.

For D2, the two gamma-ray lines of $^{60}$Co were clearly identifiable in the tagged events, but rather close together. A third line presented itself in the untagged events: the positron annihilation line at 511 keV, produced by cosmic-ray interactions in the atmosphere and balloon payload. Using these three points an in-flight energy calibration was derived and applied to all D2 PH values. The calibrated in-flight count spectrum for tagged D2 events is shown in Fig. 18, clearly showing the $^{60}$Co lines at the correct energies. The D2 count rate spectra for untagged events for the ascent and float portions of the flight are shown in Fig. 19 separately. The 511 keV line is visible on top of a power-law like continuum due to cosmic-ray interactions and internal LaBr$_3$ background. As expected, this background is higher during the ascent, since most cosmic-ray interactions take place below float altitude. Internal LaBr$_3$ lines, due to the decay of radioactive $^{138}$La, are visible at high energies (1473 keV and a broad alpha particle feature at ~2 MeV), along with a small number of untagged $^{60}$Co events.

For D1, the count spectrum for tagged singles events recorded during ground calibrations (Fig. 8) provided identifiable features that could be compared with the in-flight tagged spectrum. Specifically, we used the maximum of the combined Compton edges, the minimum of the "valley" to the low-energy side of the Compton edges, and the point at which the spectrum drops to zero on the high-energy side to perform a three-point energy calibration. The calibrated in-flight count spectrum for tagged D1 events is shown in Fig. 20; it is essentially identical to the spectrum recorded on the ground (Fig. 8).

*3.4 PSD Cuts*

The PSD cuts for identifying gamma-ray interactions in the D1 detector were determined after correcting the D1 PH values as described in Sec. 3.2. The PSD vs. corrected D1 PH distribution for flight data is shown in Fig. 21. By comparing these data with the ground calibration performed using a $^{252}$Cf source (Fig. 9), we defined a cut (gray line in Fig. 21) between the neutron and gamma-ray tracks in the two-dimensional data space. All events whose (D1 PH, PSD) values fall below the gray line in Fig. 21 were considered gamma-ray interactions in D1.

*3.5 ToF Resolution for Tagged Events*

Once the corrections described above (temperature effects, energy calibration, and PSD cuts) had been derived, it was possible to apply them to coincident events and evaluate the performance of SolCompT as a Compton telescope while in flight. The ToF resolution was measured using tagged $^{60}$Co coincident events. All tagged events from the entire flight were included which passed the following cuts: (D1 PH, PSD) combination indicated a gamma-ray interaction; $E_1 \geq$ 75 keV; $E_2 \geq$ 75 keV; $E_{tot} \leq$ 2 MeV; and Compton scatter angle $\varphi \leq$ 90º (see Eq. 1). The ToF histogram for these events was well represented by a narrow Gaussian. We assigned the centroid of this Gaussian to a ToF value of +0.5 ns, as appropriate for a downward-scattered gamma ray. The final SolCompT in-flight ToF spectrum for tagged events is shown in Fig. 22. This plot includes events with total energy between ~200 keV and 2 MeV, though most events are concentrated between ~500 keV and ~1.5 MeV. The ToF resolution is 760 ± 30 ps (FWHM). SolCompT has thus demonstrated sub-ns ToF resolution under balloon flight conditions, as required for future scintillator-based Compton telescopes with SiPM readouts.

*3.6 Energy Spectrum for Tagged Events*

The directionality and energy response of the SolCompT instrument in flight could also be studied using tagged $^{60}$Co coincident events, following the basic procedure used to produce the $^{22}$Na spectrum during ground calibrations (Sec. 2.4). We selected tagged gamma-ray events using the same energy cuts as in Sec. 3.5, but with an additional requirement of 0 ns ≤ ToF ≤ 1.5 ns (i.e., downward-moving gamma rays) and no restriction on the Compton scatter angle. The computed scatter angle $\varphi$ is plotted vs. $E_{tot}$ in Fig. 23. Although there are only a few hundred points, two clusters can be identified at the correct energies for $^{60}$Co and near a scatter angle of ~45º, as expected for the position of the tagged source relative to the instrument axis (Sec. 2.2). These points represent completely absorbed Compton scatter events, reconstructed at the correct energy and scatter angle.

We next imposed a scatter angle cut of 30º ≤ $\varphi$ ≤ 55º (dashed lines in Fig. 23) and formed an energy count spectrum from these correctly-reconstructed events (Fig. 24). The statistics are quite poor (only 185 events) as expected, due to the small size of the SolCompT instrument and the short duration of the balloon flight. Nevertheless, the two gamma-ray lines of $^{60}$Co are clearly present at the correct energies. The poor statistics prevent a detailed analysis of the energy resolution; nevertheless, Figs. 23 & 24 demonstrate qualitatively that the SolCompT instrument operated correctly as a Compton telescope during flight in terms of its angular and energy response.

*3.7 ToF Spectrum for Untagged Events*

Using the same calibration and corrections described in Sec. 3.5, we examined the ToF distribution for untagged events. The ToF spectrum for untagged events accumulated at float with 800 keV ≤ $E_{tot}$ ≤ 2 MeV is shown in Fig. 25. Ideally this spectrum would show two clearly separated peaks, corresponding to downward-moving (positive ToF) and upward-moving (negative ToF) double-scatter gamma rays, as was the case for the FACTEL balloon experiment [32]. This clean ToF separation was enabled by the fact that the distance between the D1 and D2 layers in FACTEL was ~30 cm, and the instrument was flown at the top of the balloon gondola with very little surrounding mass. In contrast, in SolCompT the D1 - D2 separation was only 15 cm, and the instrument was surrounded on all sides by CSBF equipment (Fig. 13). Thus the untagged ToF spectrum seen in Fig. 25 is more difficult to interpret: based on the tagged ToF spectrum (Fig. 22) the downward- and upward-moving peaks are expected to overlap slightly, and they are likely to sit on top of a broad distribution centered near ~0 ns. This broad component, studied in detail in the COMPTEL background [39], is the result of cosmic-ray interactions in passive material surrounding the instrument that give rise to multiple-photon radioactive decays and/or showers of secondary particles.

We have attempted to interpret the SolCompT untagged ToF spectrum (Fig. 25) as the superposition of three Gaussians representing the components described above. Two narrow Gaussians were fixed at centroid values of +0.5 ns (dotted line) and -0.5 ns (dashed line) to represent downward- and upward-moving gamma rays, respectively. The widths of these Gaussians were also fixed to match the measured value for tagged gamma rays (760 ps FWHM; Fig. 22). A third, broad Gaussian centered on ~0 ns (dot-dashed line) was included to represent the locally generated background. The fit shown in Fig. 25 indicates that this model is indeed a plausible interpretation of the data, given the measured ToF spectrum for tagged events (Fig. 22) and the physical understanding gained by the experience with COMPTEL [39] and FACTEL [32]. The simple model is clearly not quantitatively correct; however, since a full understanding would require detailed Monte Carlo simulations of background generation in the CSBF test gondola, which is beyond the scope of this work, we have not attempted to model the untagged ToF spectrum any further. The results obtained using tagged $^{60}$Co events demonstrate successful operation of the instrument, which was the goal of the SolCompT balloon flight.

## 4. Conclusions and Future Plans

The SolCompT balloon experiment has successfully demonstrated stable in-flight performance of a Compton telescope prototype consisting of advanced scintillator detector elements with SiPM readouts. Custom readout electronics produced fast signals and linear PH response despite the capacitance of the large-area SiPM arrays, and custom bias voltage supplies kept the gain approximately stable despite significant temperature variations. Residual (~10%) gain variations were easily corrected using singles events from the tagged $^{60}$Co source. Using tagged coincident events over a wide energy range of 200 keV $\leq E_{tot} \leq$ 2 MeV, the in-flight ToF resolution was measured to be 760 ± 30 ps (FWHM). Proper angular and energy response for Compton events were both apparent, although statistics were poor. Ground calibrations demonstrated energy resolution of ~5% (FWHM) above 1 MeV. The SolCompT experiment has therefore demonstrated that scintillator-based Compton telescopes with SiPM readouts hold great promise for future gamma-ray astronomy and solar physics instrumentation.

We have begun the development of a larger balloon-borne Compton telescope at UNH as part of the Advanced Scintillator Compton Telescope (ASCOT) project [34]. This instrument will employ arrays of organic p-terphenyl scintillator in D1 and inorganic CeBr$_3$ scintillator in D2, read out by MicroFC-60035-SMT SiPMs from SensL Technologies Ltd [40]. Compared to the Hamamatsu devices used in SolCompT, the newer SensL SiPMs have the advantages of a dedicated "fast" output for improved ToF performance, a lower bias voltage of ~30 V, and reduced temperature sensitivity (~22 mV/ºC). The ASCOT instrument will fly on a CSBF one-day turnaround balloon flight, without a pressure vessel, in order to image the Crab Nebula at ~MeV energies and fully space-qualify our advanced scintillator/SiPM detector technology. The ASCOT balloon flight is currently scheduled for September of 2017.

## Acknowledgments

We would like to thank the staff of NASA's Columbia Scientific Balloon Facility for their invaluable support preparing and conducting the SolCompT balloon flight. This work was supported by NASA grants NNX12AB36G and NNX13AC89G.

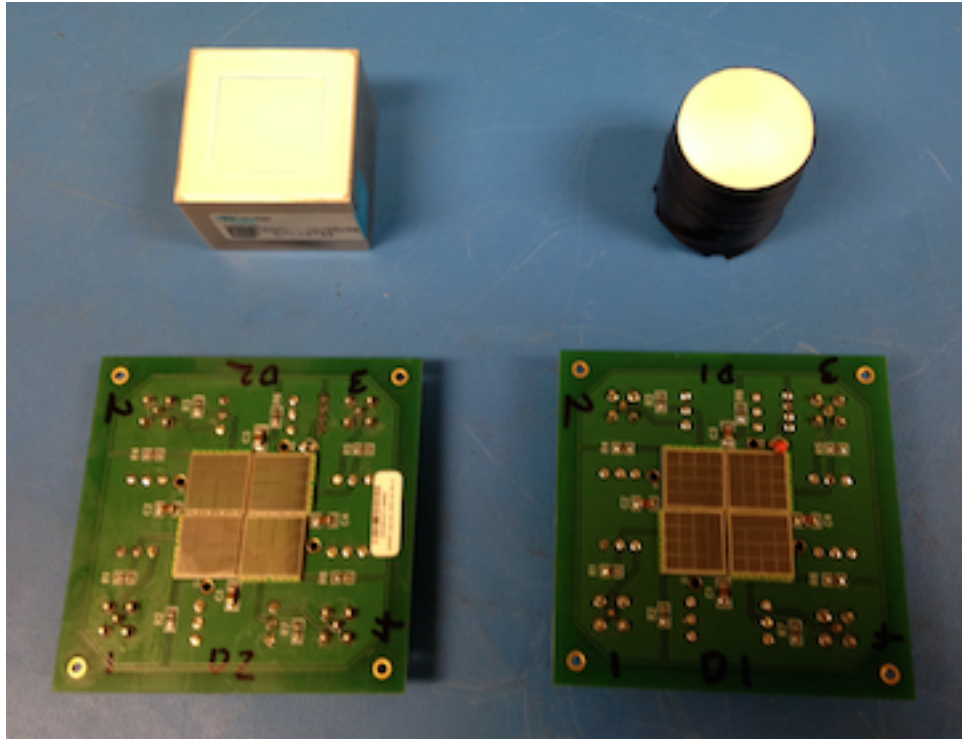

Fig. 1. The SolCompT D1 stilbene crystal (right), D2 LaBr$_3$ crystal (left), and Hamamatsu S11828-3344 SiPM readout arrays (bottom).

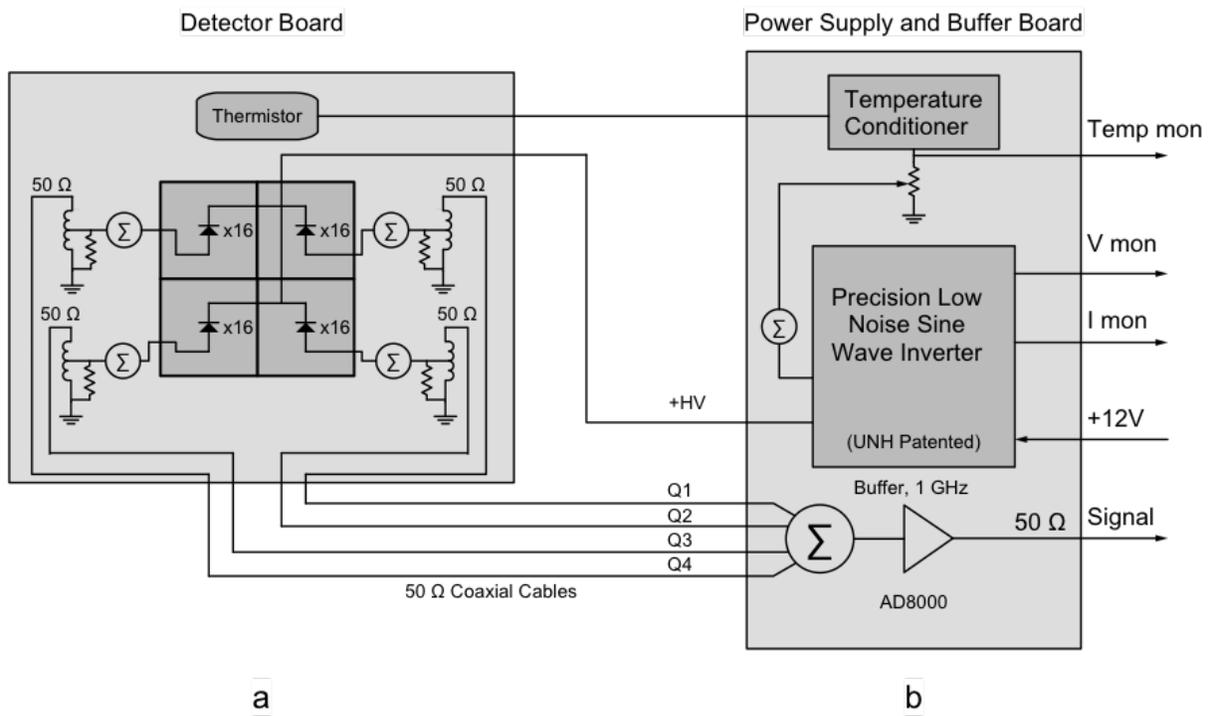

Fig. 2. Block diagram of the SolCompT front-end electronics, including (a) detector board, and (b) power supply/buffer board. See text for details.

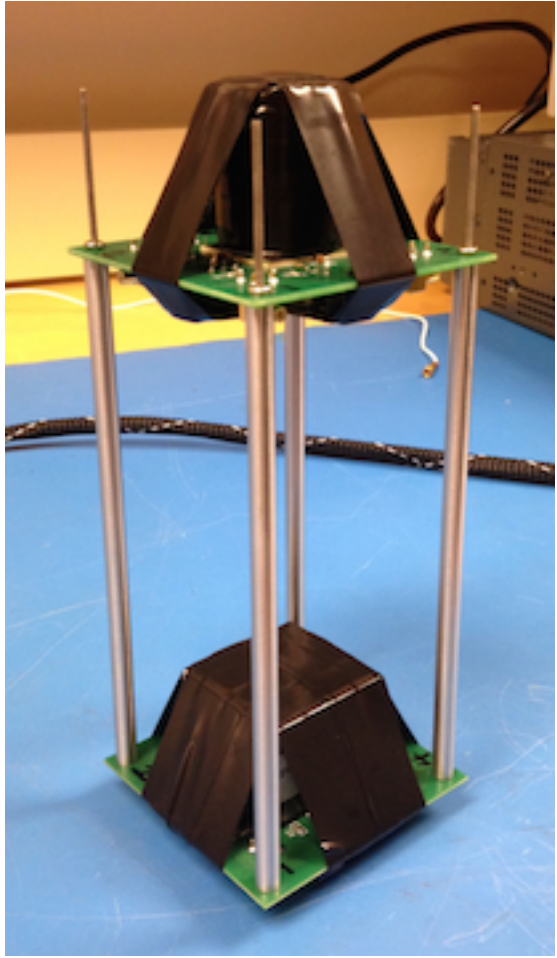

Fig. 3. Assembled SolCompT two-element Compton telescope, with 15 cm center-to-center scintillator separation.

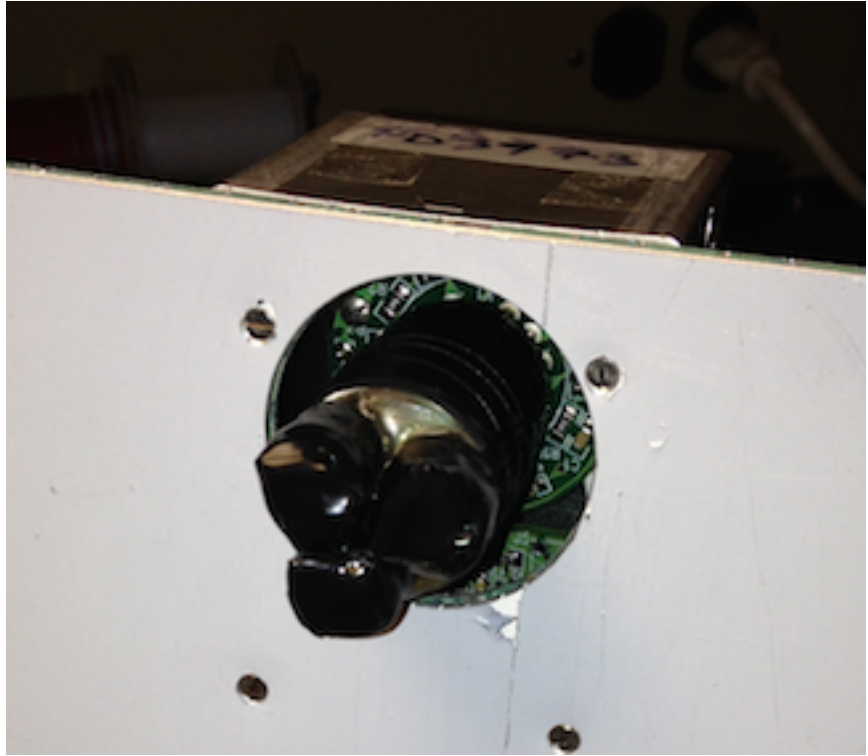

Fig. 4. The tagged calibration source for the SolCompT instrument, showing three ~80 nCi cylinders of $^{60}$CO embedded in plastic scintillator, wrapped in black electrical tape, and coupled to the face of an R1924A PMT with custom front-end electronics.

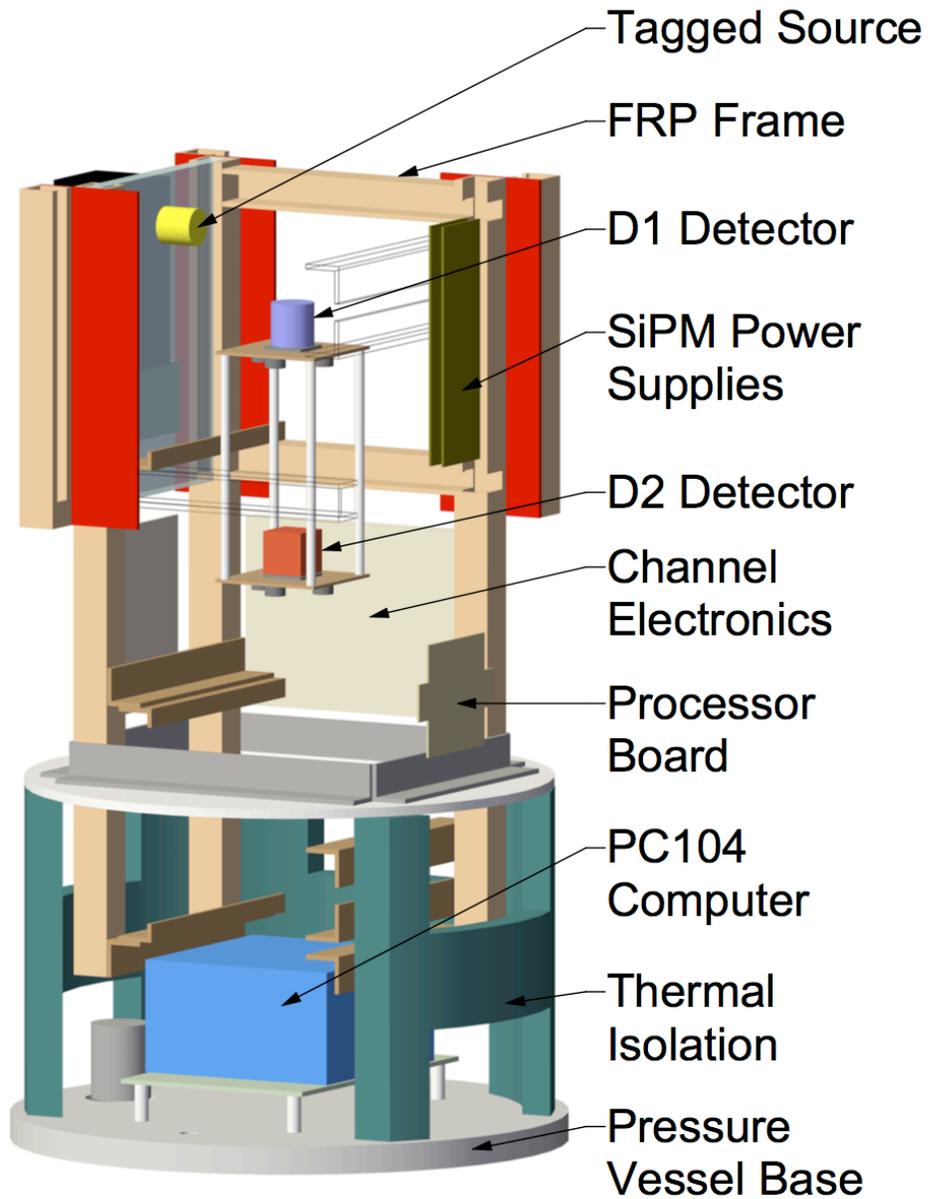

Fig. 5. Cutaway drawing of the SolCompT balloon payload with the major components labeled. See text for detailed descriptions.

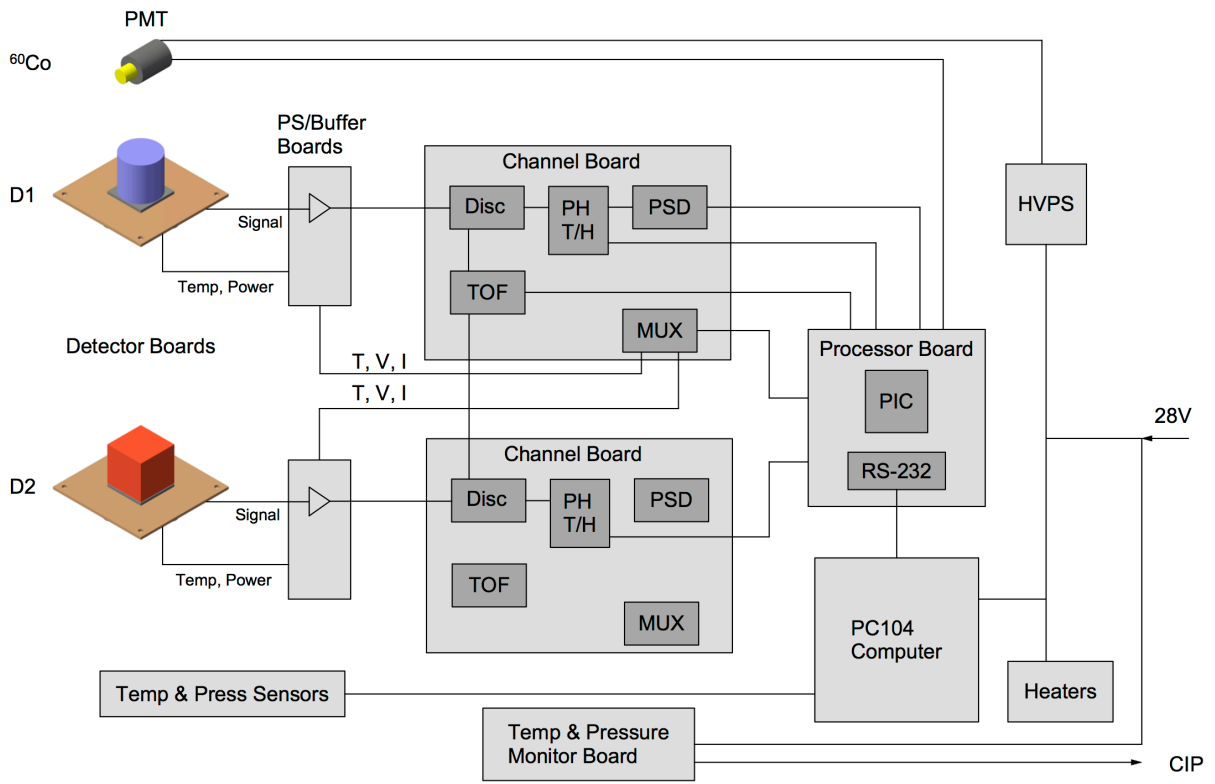

Fig. 6. System block diagram of the SolCompT balloon payload. See text for details.

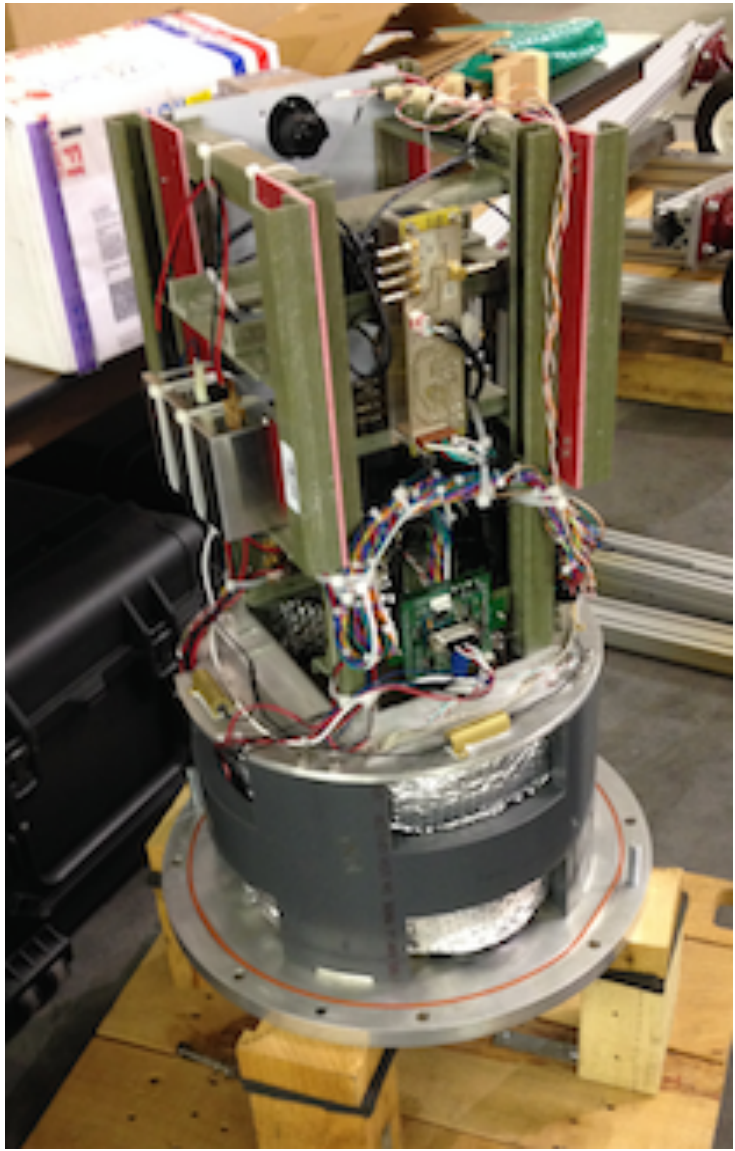

Fig. 7. The assembled SolCompT balloon payload as flown, shown post-flight in the field (Fort Sumner, NM) with the pressure vessel cover removed.

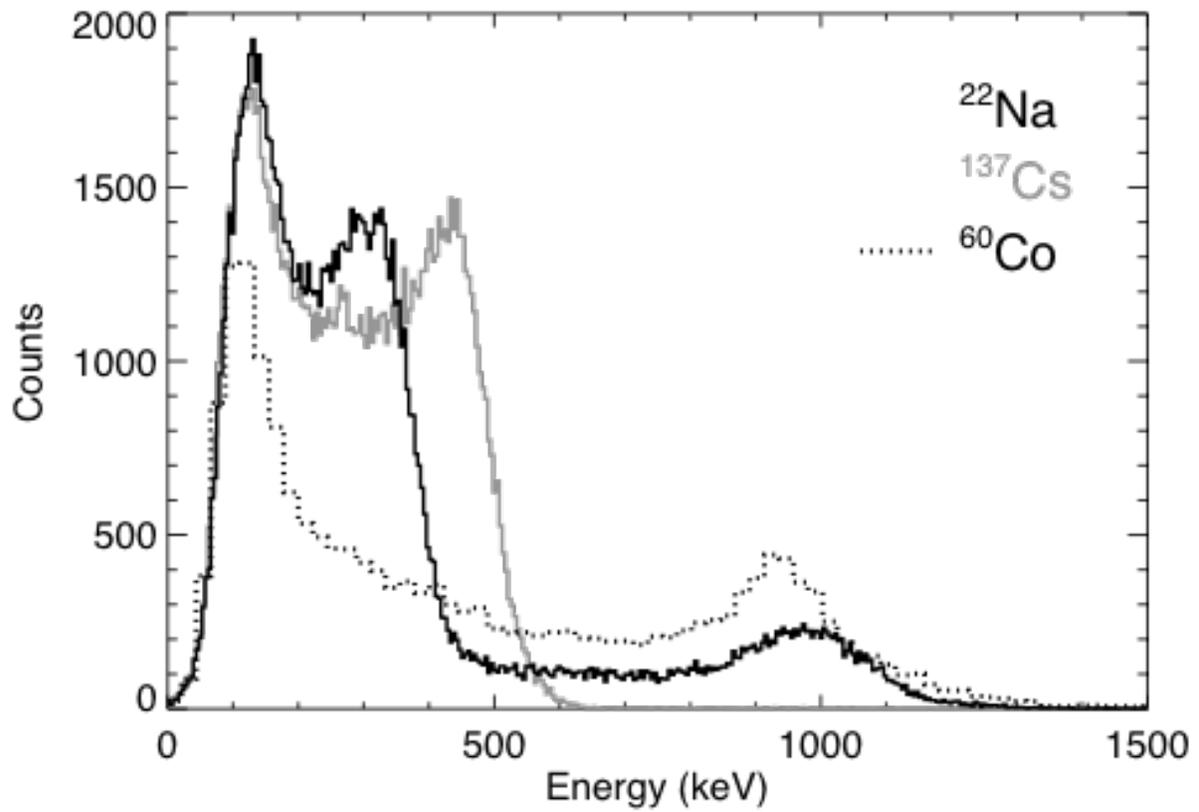

Fig. 8. Calibrated singles spectra recorded by the D1 stilbene detector, including $^{22}$Na, $^{137}$Cs, and the tagged $^{60}$Co source.

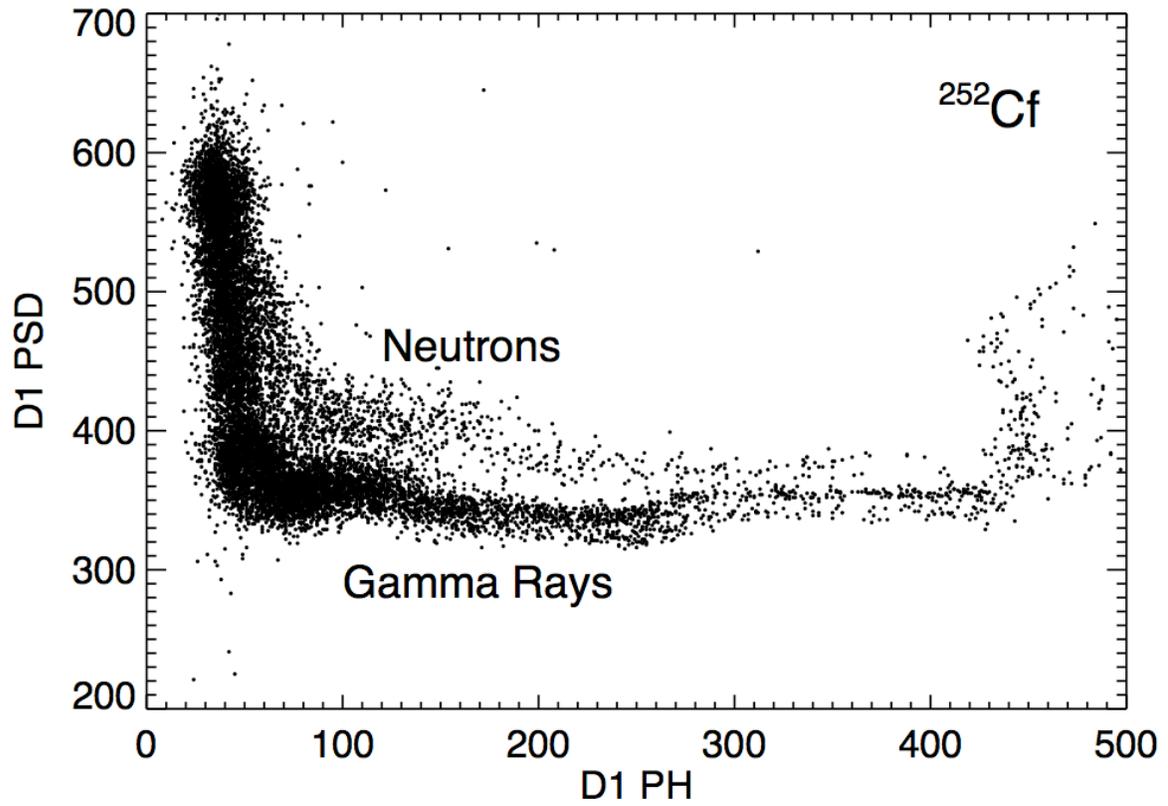

Fig. 9. PSD value vs. PH for the D1 detector, measured for a $^{252}$Cf fission source. Two tracks can be seen corresponding to neutron (upper) and gamma-ray (lower) interactions.

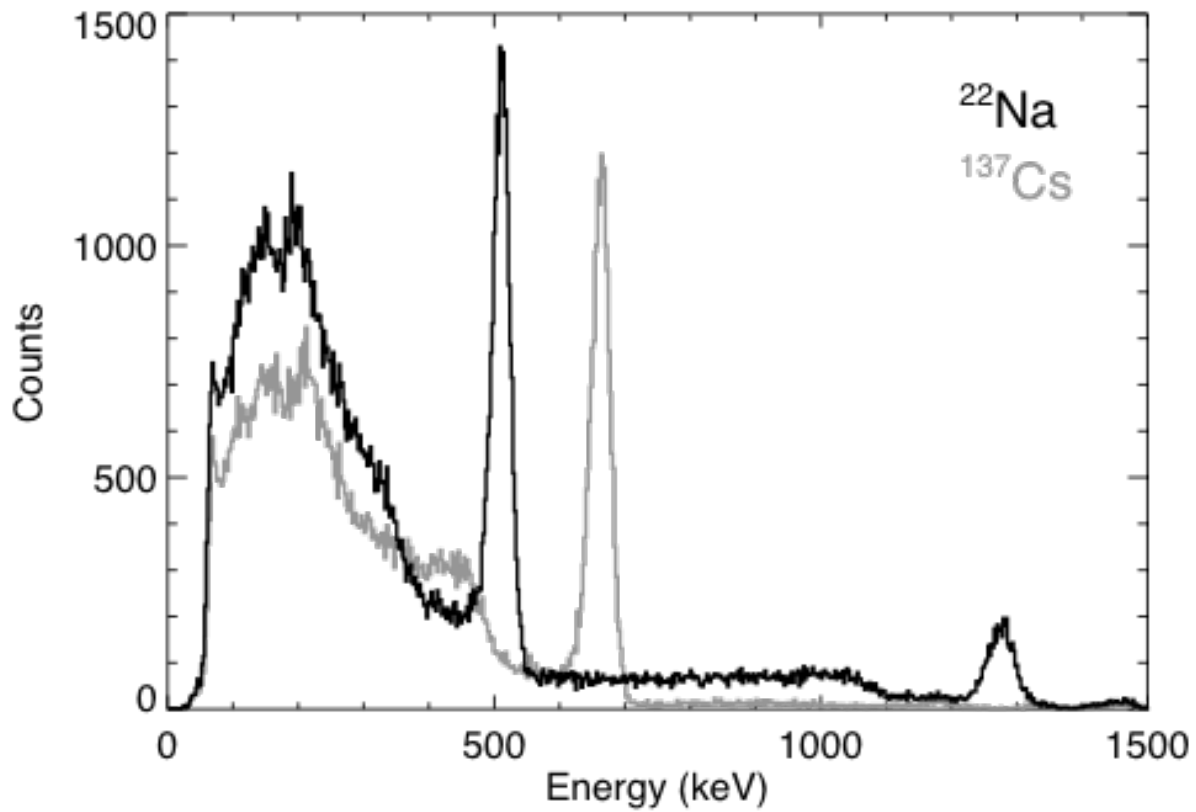

Fig. 10. Calibrated singles spectra recorded by the D2 LaBr$_3$ detector for $^{22}$Na and $^{137}$Cs. The energy resolution at 662 keV is 5.30 ± 0.04% (FWHM).

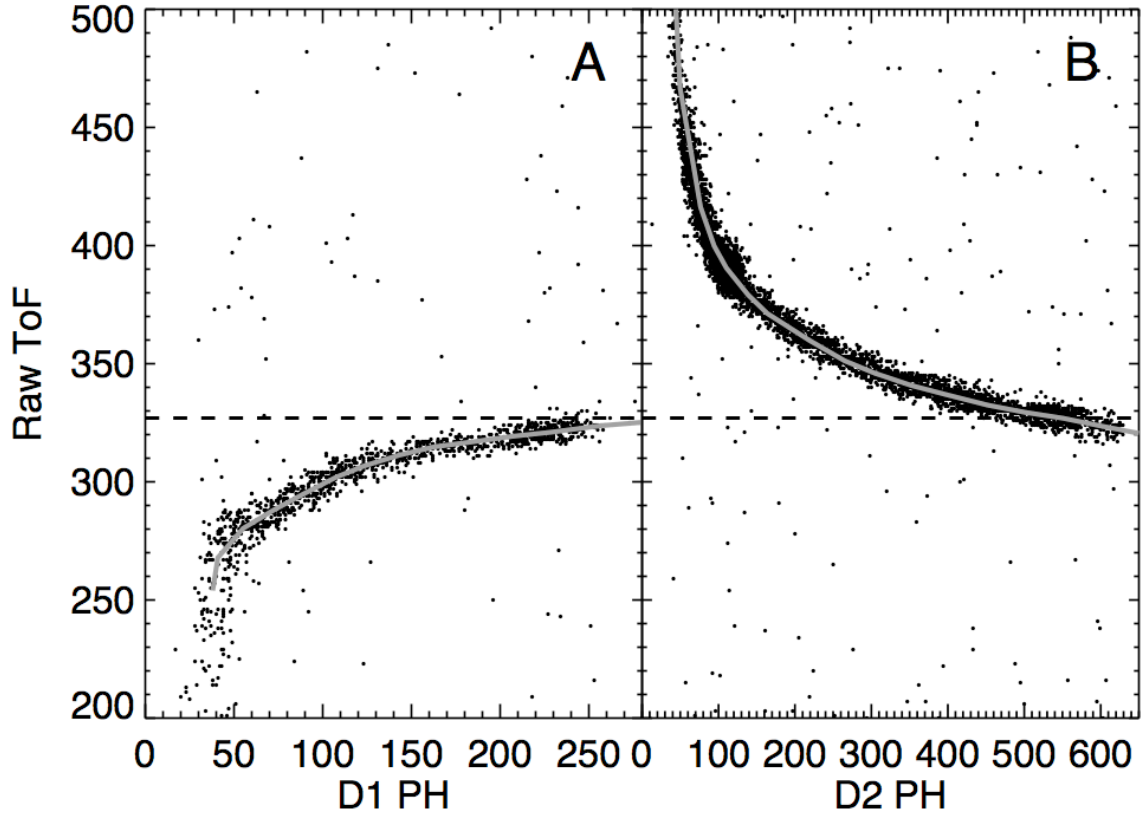

Fig. 11. ToF offset (gray line) derived as a function of PH individually for D1 (a) and D2 (b) relative to an arbitrary value (ToF = 327; dashed line), measured using coincident events from a bright $^{60}$Co source placed between the two detectors. See text for details.

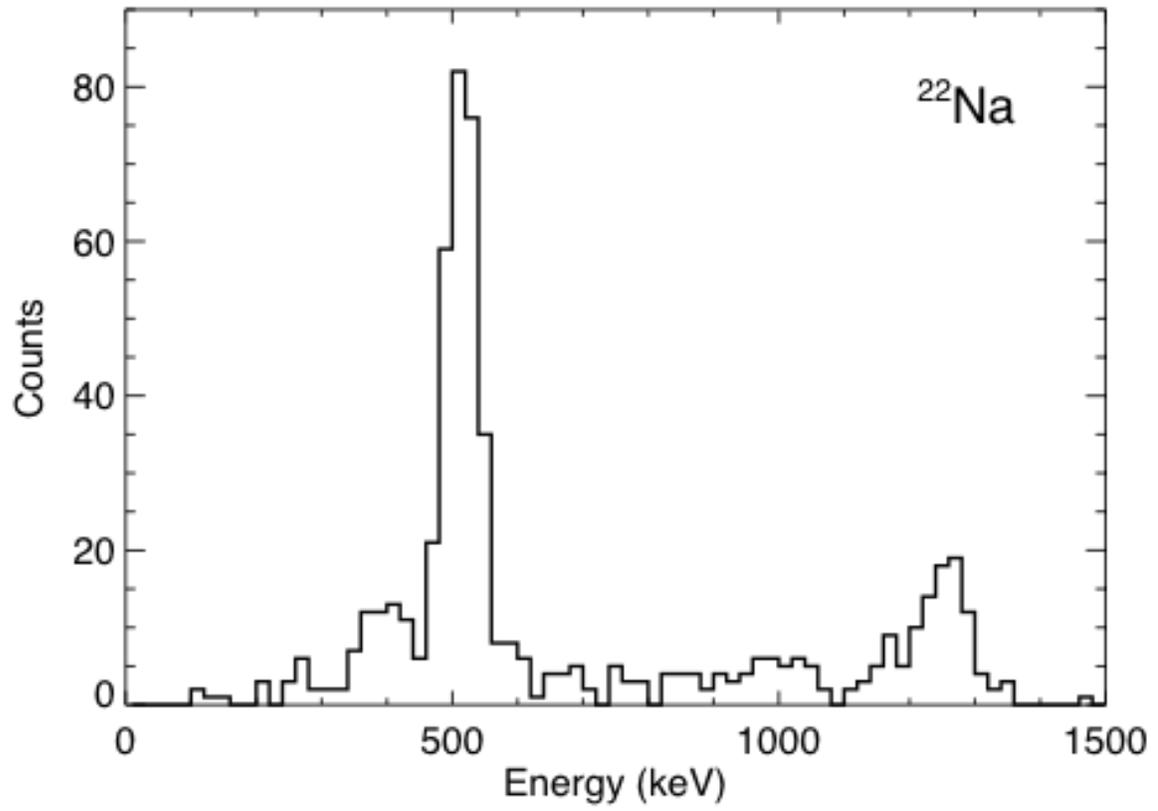

Fig. 12. Energy spectrum of Compton scatter events recorded from a $^{22}$Na source placed ~40º off the telescope axis, after applying ToF and $\varphi$ cuts.

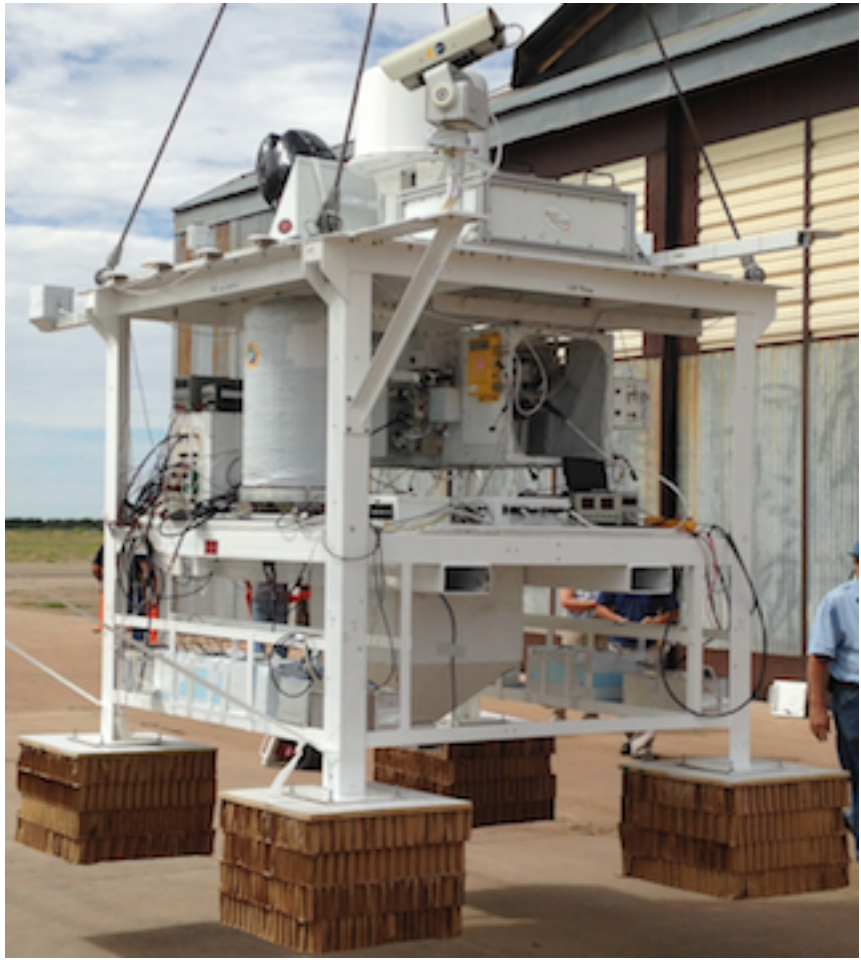

Fig. 13. The SolCompT payload (white cylinder left of center) integrated into a CSBF balloon gondola during pre-flight testing in Fort Sumner, NM.

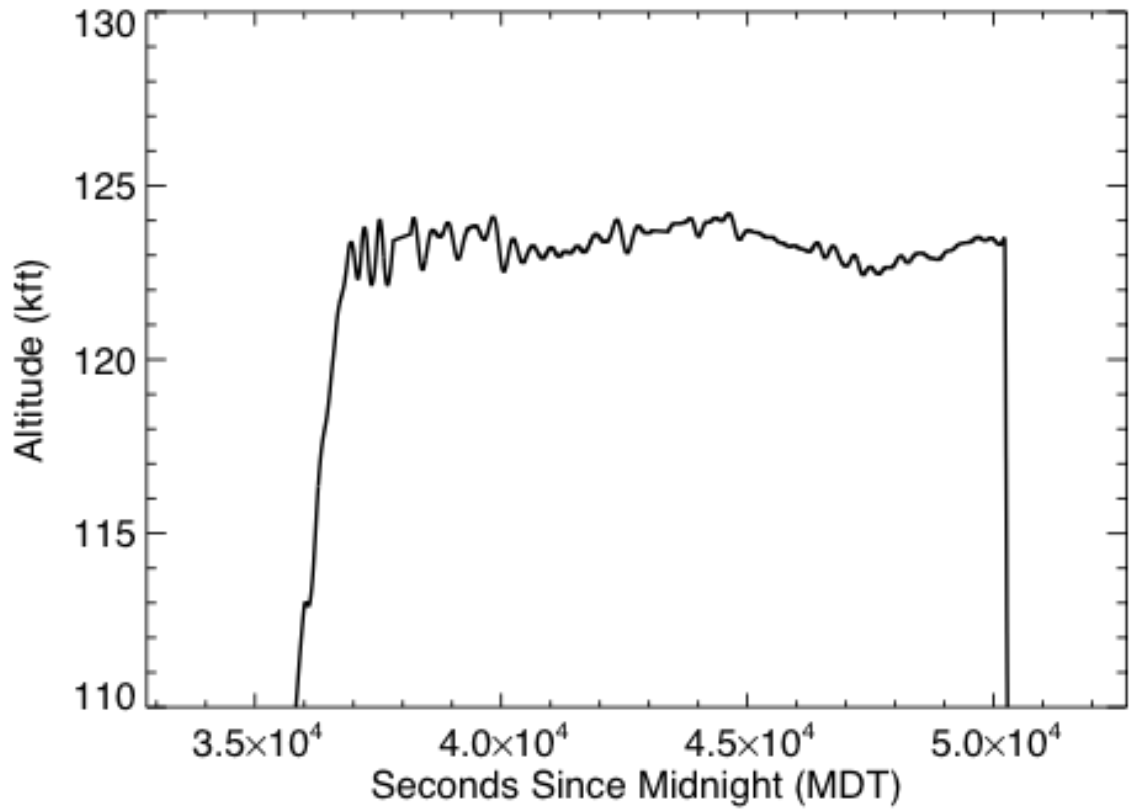

Fig. 14. Altitude profile of the CSBF LDB test balloon flight carrying SolCompT. Time is in seconds since local midnight (Mountain Daylight Time). The payload maintained a float altitude of ~123,000 feet for ~3.75 hours.

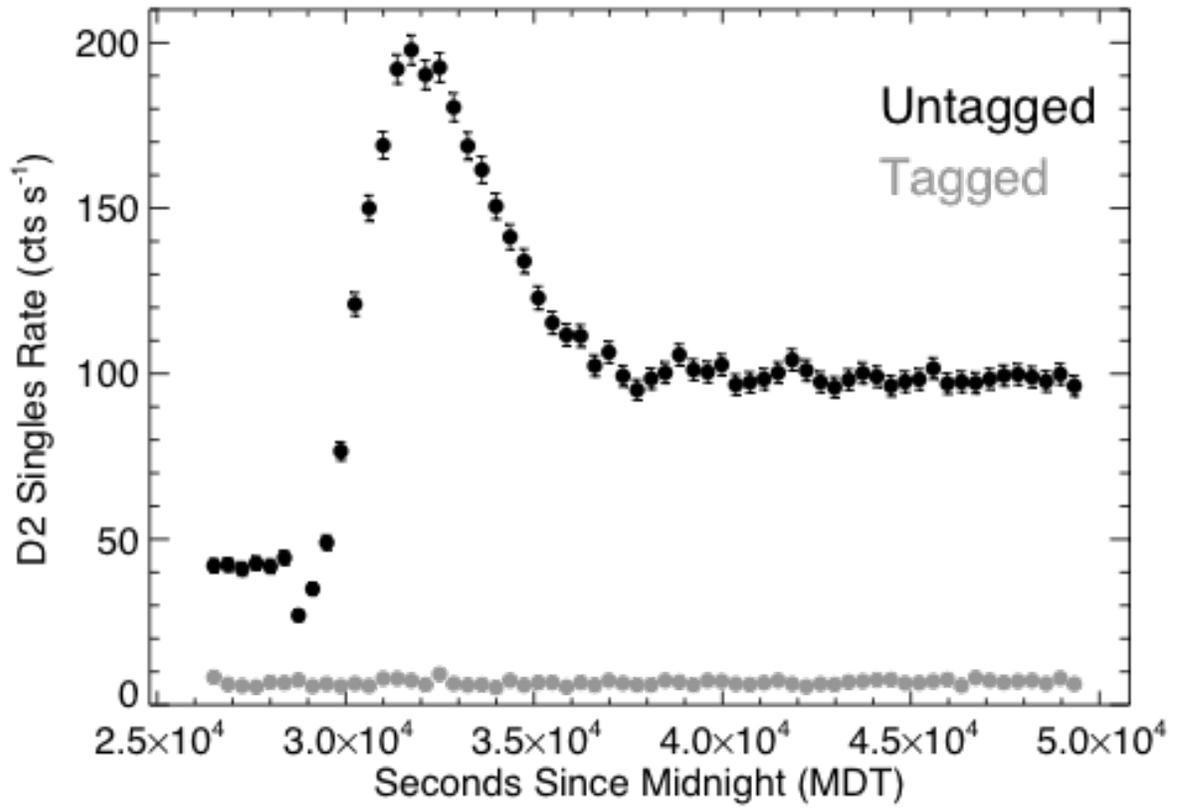

Fig. 15. Singles count rate in the SolCompT D2 detector for both untagged (black) and tagged (gray) events. The untagged events show the expected behavior (see text), while the tagged event rate is constant throughout the flight.

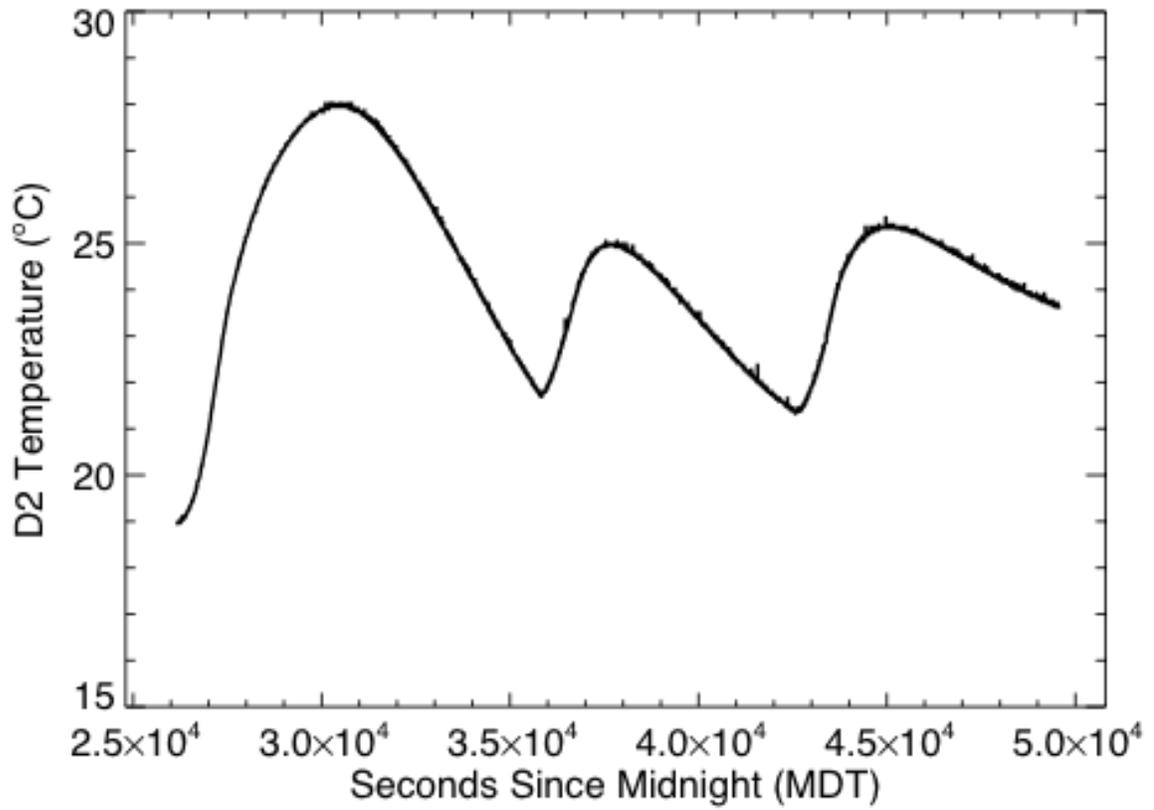

Fig. 16. Temperature vs. time recorded by the thermistor bead mounted on the D2 detector board during the SolCompT balloon flight. Variations at times after ~$3.5 \times 10^4$ s are due to the heaters being cycled on and off.

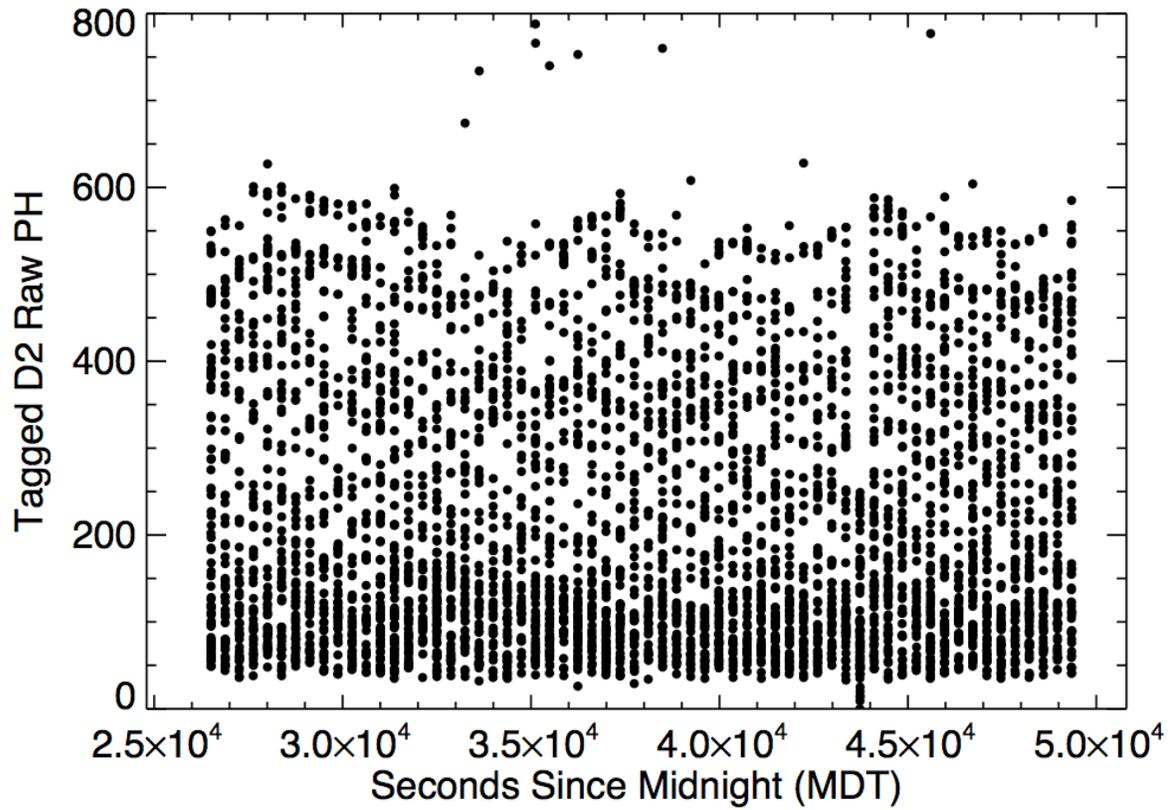

Fig. 17. Raw PH vs. time recorded for tagged events by the SolCompT D2 detector during the balloon flight. Although the gain is held approximately constant, residual variations of ~10% are apparent due to temperature changes.

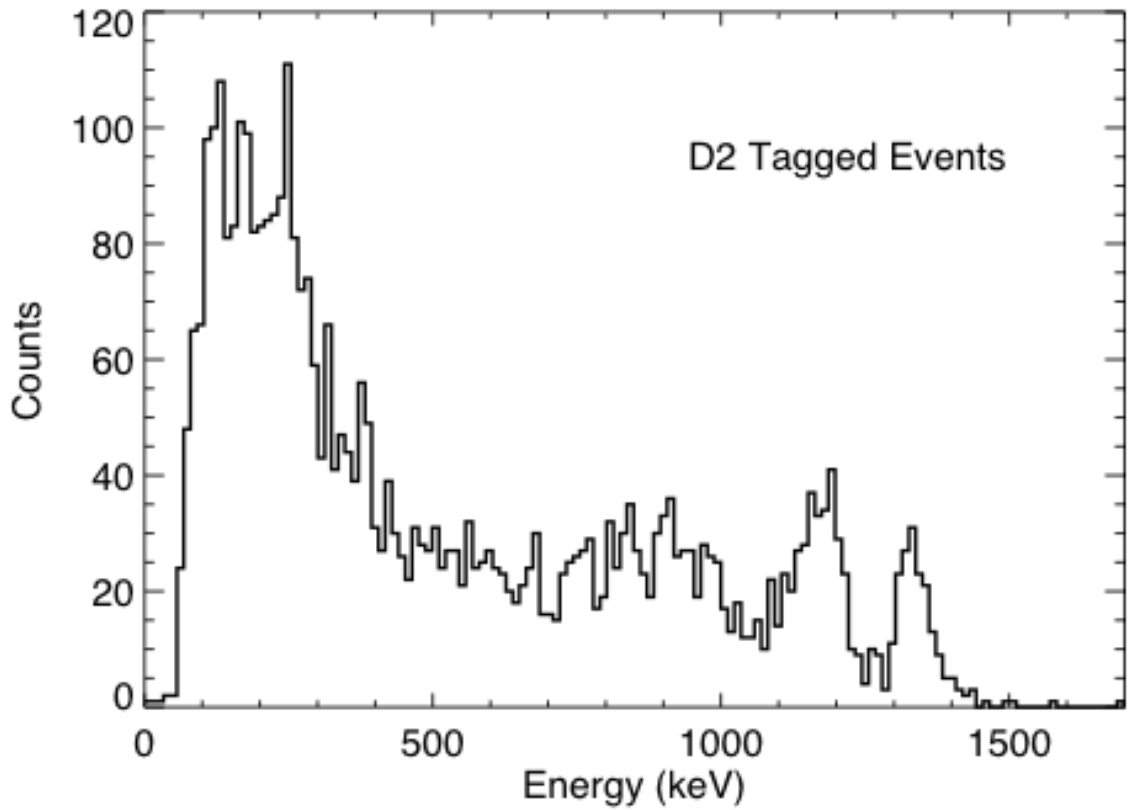

Fig. 18. Calibrated in-flight energy count spectrum of tagged D2 singles events. The two gamma-ray lines of $^{60}$Co are clearly visible at 1173 keV and 1333 keV.

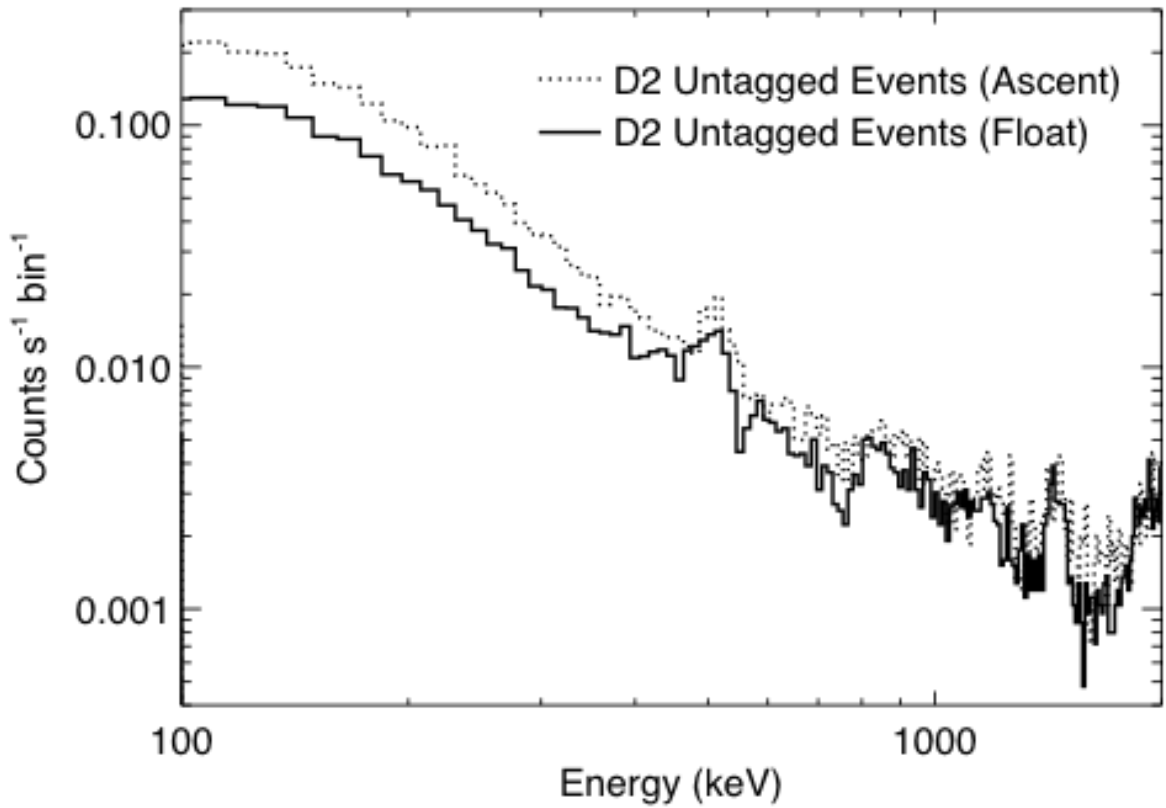

Fig. 19. Calibrated in-flight energy count rate spectra of untagged D2 singles events. Events accumulated during the ascent (dotted) and at float (solid) are plotted separately. The recorded events are a combination of cosmic ray induced background in the atmosphere and balloon payload, including the line at 511 keV, and internal $LaBr_3$ background due to $^{138}La$ decay.

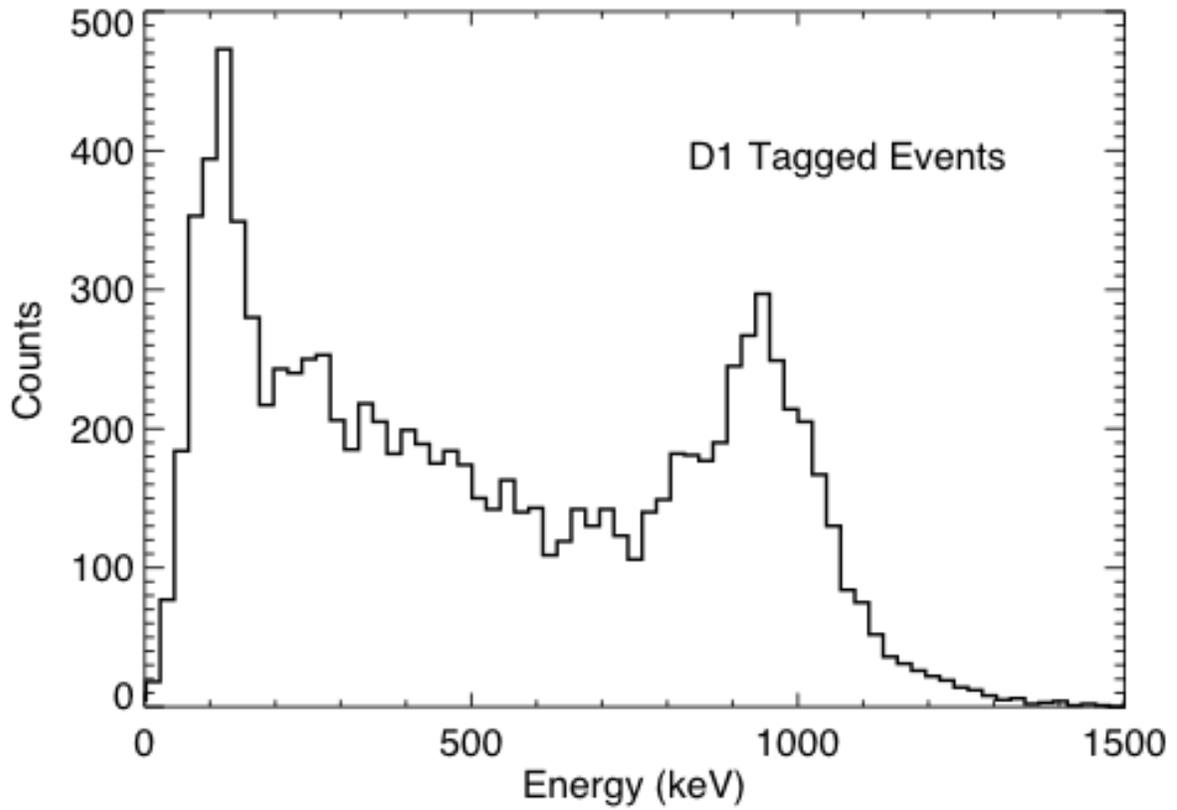

Fig. 20. Calibrated in-flight energy count spectrum of tagged D1 singles events. The combined Compton edges of the $^{60}$Co lines form the prominent peak at ~950 keV.

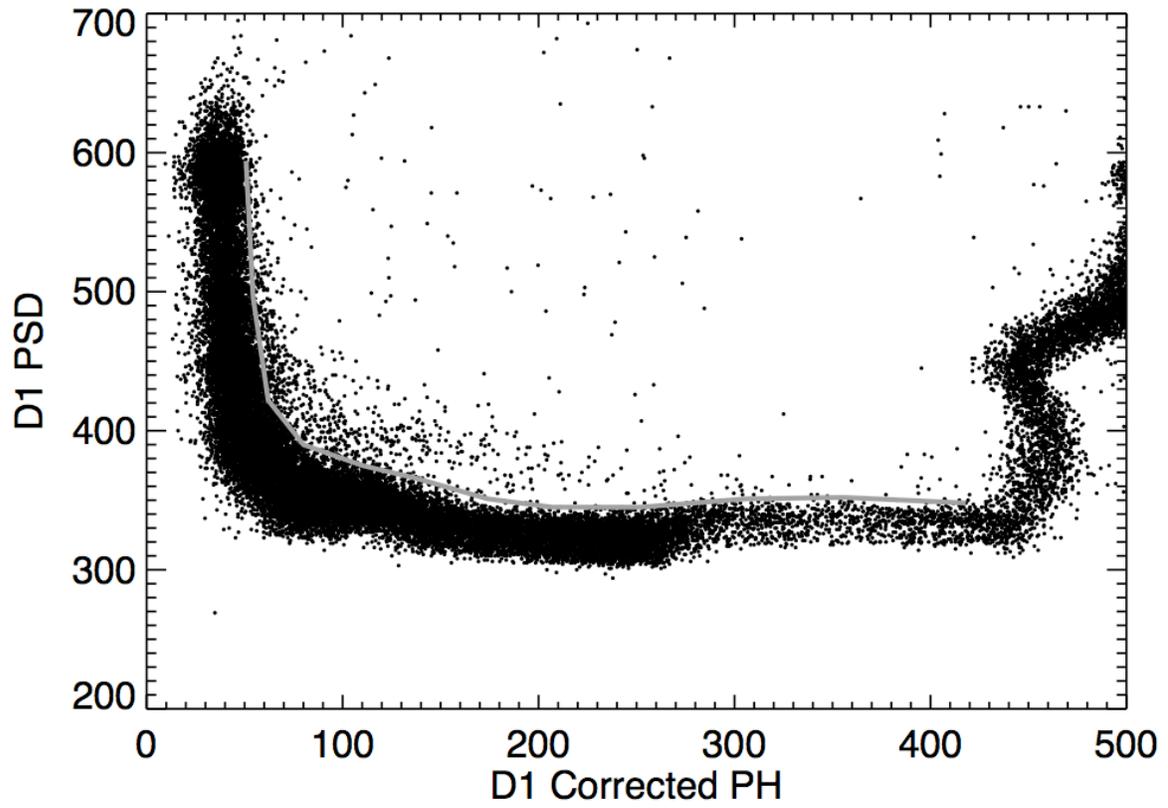

Fig. 21. PSD value vs. corrected D1 PH for flight singles data. The gray line separates the neutron interactions (above) from gamma-ray interactions (below).

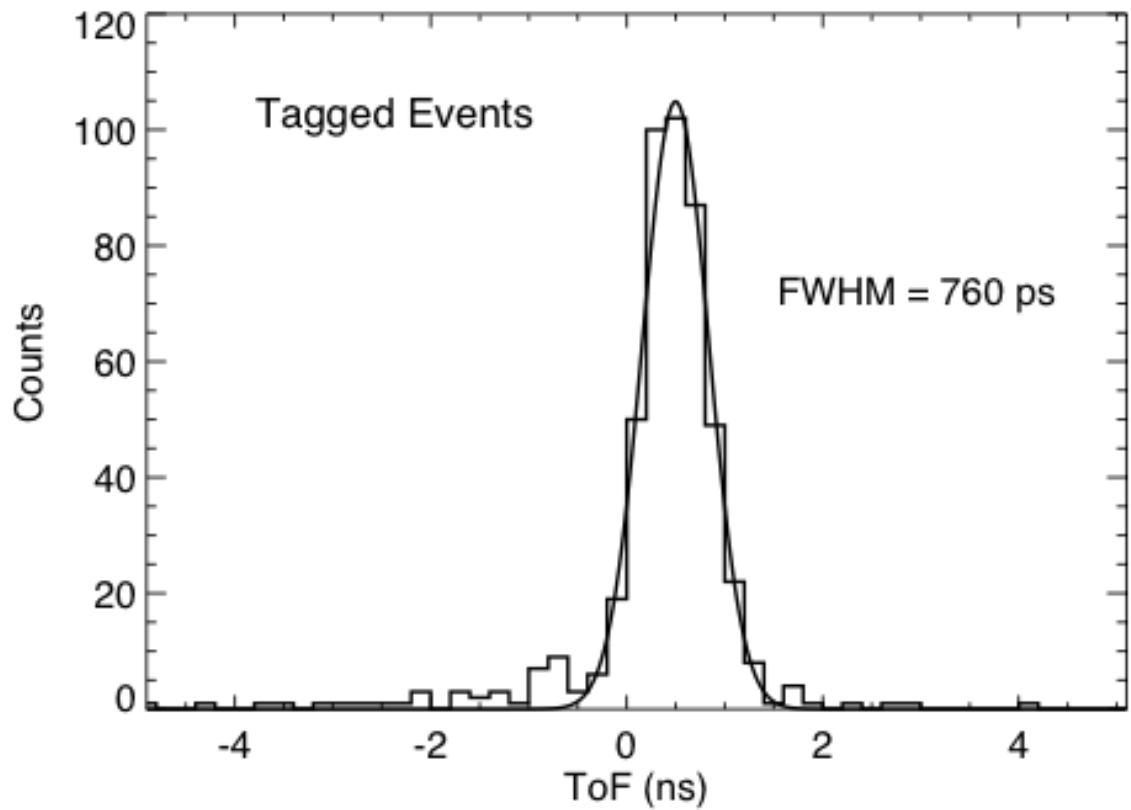

Fig. 22. In-flight SolCompT ToF spectrum measured using the tagged $^{60}$Co source. Tagged coincident events from the entire flight were included if they passed basic data cuts (see text). The ToF resolution is $760 \pm 30$ ps (FWHM).

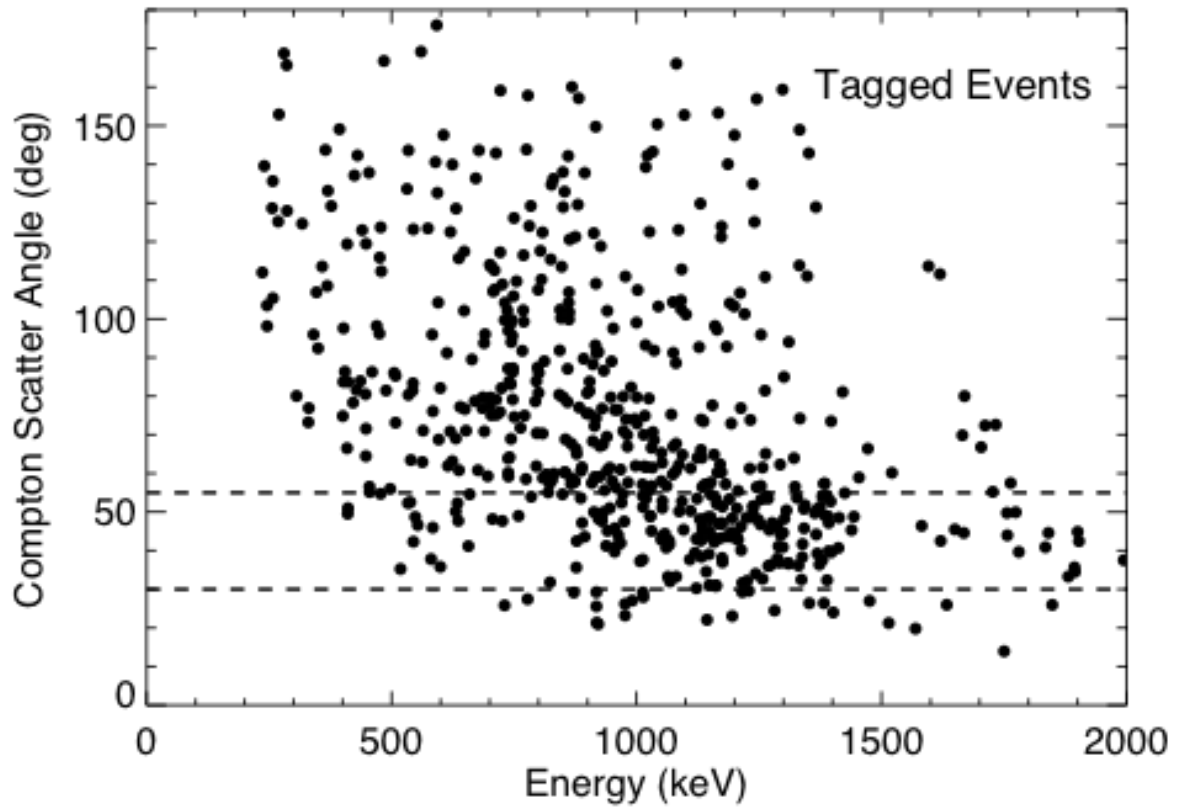

Fig. 23. Calculated Compton scatter angle $\varphi$ vs. $E_{tot}$ for tagged coincident events during the SolCompT balloon flight. Correctly reconstructed events lie with a scatter angle range of approximately 30º ≤ $\varphi$ ≤ 55º (dashed lines). The two gamma-ray lines of $^{60}$Co are visible within this range.

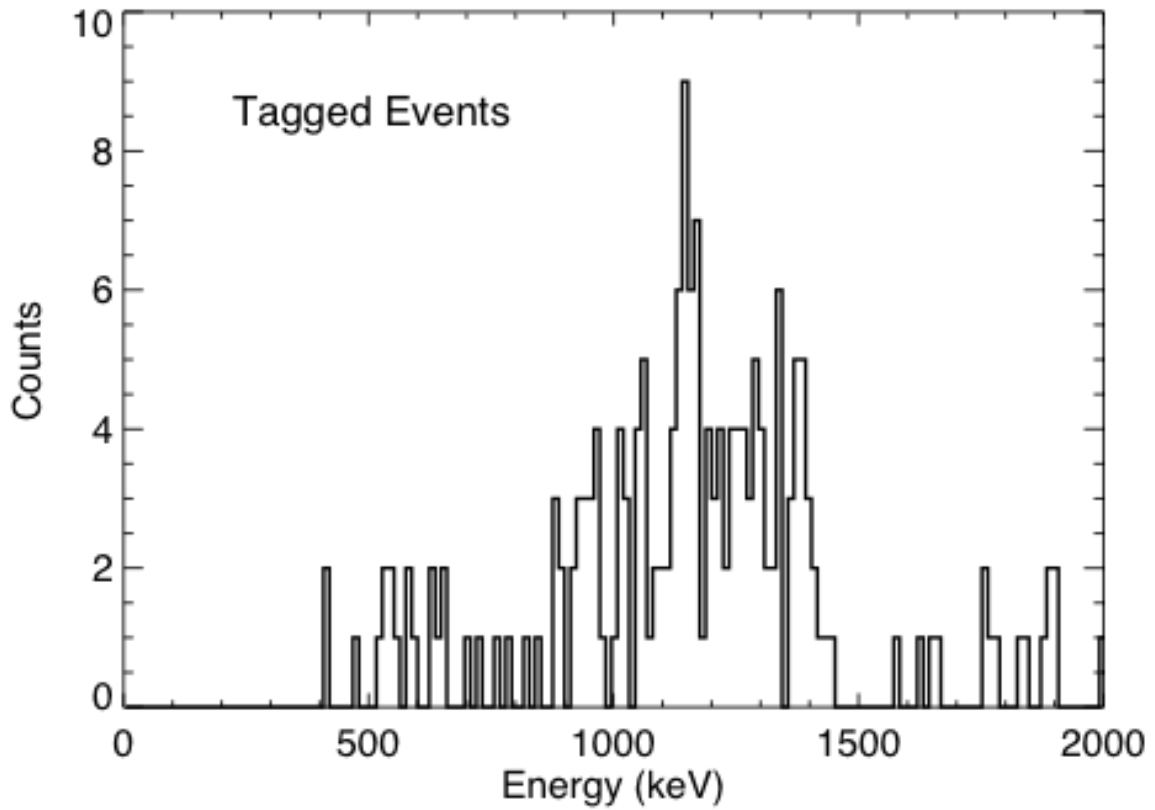

Fig. 24. Energy count spectrum of the $^{60}$Co source for correctly reconstructed tagged coincident events ($30° \leq \varphi \leq 55°$). The two gamma-ray lines are present at approximately the correct energies, although the statistics are too poor to permit detailed analysis.

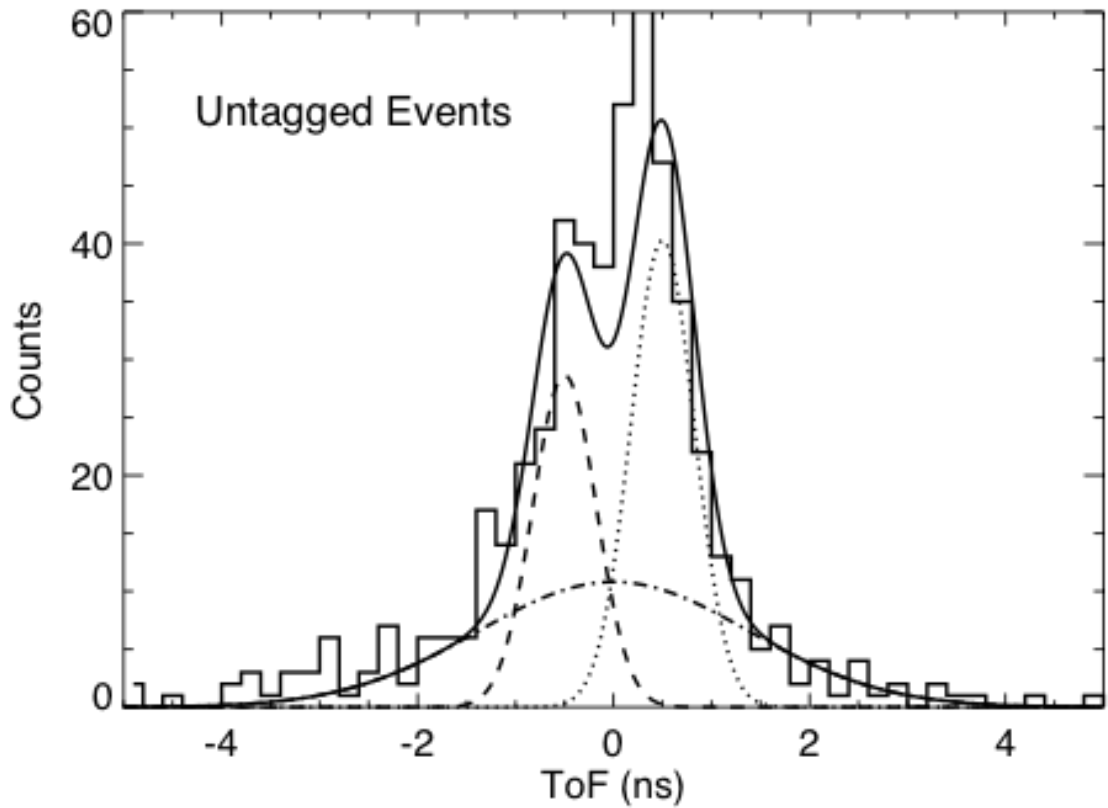

Fig. 25. In-flight SolCompT ToF spectrum for untagged events with 800 keV $\leq E_{tot} \leq$ 2 MeV measured at float. The spectrum is fit with a simple model consisting of downward-moving gamma rays (dotted line), upward-moving gamma rays (dashed), and locally generated background (dot-dashed).